\documentclass{CSML-by}
\usepackage{lastpage}
\def\dOi{13(2:16)2017}
\lmcsheading%
{\dOi}
{1--\pageref{LastPage}}
{}
{}
{May\phantom..~\phantom06, 2014}
{Jun.~30, 2017}
{}

\newif\ifdraft\draftfalse 
\newif\iftr\trfalse       
\newif\ifjournal\journalfalse 
\newif\ifieee\ieeefalse   
\newif\ifattic\atticfalse  

\journaltrue

\newif\iffull\fullfalse\iftr\fulltrue\fi\ifjournal\fulltrue\fi
\newif\ifconf\conftrue\iffull\conffalse\fi
\newif\ifspringer\springertrue\ifieee\springerfalse\fi
\ifjournal\trfalse\fi

\ifjournal

\usepackage{natbib}
\usepackage{hyperref}\hypersetup{hidelinks}
\usepackage{bm}

\theoremstyle{plain}

\usepackage{amssymb,amsfonts,latexsym,alltt,xspace}
\else
\ifieee 

\documentclass[nocopyrightspace,\iffull,nocopyrightspace\fi,natbib]{sigplanconf}  
\usepackage{amssymb}
\usepackage{amsthm}

\usepackage{url}
\usepackage{xspace}
\bibpunct();A{},
\let\cite=\citep

\iftr\usepackage{msrtr-cam}\fi

\else 
\documentclass{llncs}
\usepackage{amssymb}
\usepackage{url}
\usepackage{natbib}
\bibpunct();A{},

\usepackage{xspace}
\iftr\usepackage{msrtr-cam}\fi
\fi\fi

\usepackage{mathptmx}

\usepackage{amsfonts, amsmath, amssymb, stmaryrd}
\usepackage{listings}
\usepackage{xspace}

\usepackage{algorithm}
\usepackage{algorithmic}
\usepackage{amsfonts}
\usepackage{amsmath}
\usepackage{amssymb}
\usepackage{multirow}
\usepackage{xspace}
\usepackage{url}
\usepackage{multirow}
\usepackage{graphicx}

\usepackage{color}
\definecolor{dkblue}{rgb}{0,0.1,0.5}
\definecolor{dkgreen}{rgb}{0,0.4,0}
\definecolor{dkred}{rgb}{0.6,0,0}
\definecolor{linkColor}{rgb}{0,0,0.5}

\newcommand\maybecolor[1]{\color{#1}}

\ifdraft

\def\Black{\color{black}}

\else
\def\Black{}
\ifconf   
  \renewcommand{\color}[1]{}
\fi
\fi

\usepackage{url}
\usepackage{listings}
\usepackage{version}
\usepackage{extarrows}

\newenvironment{prog}{\begin{array}[t]{@{}l@{}}}{\end{array}}

\def\condarray#1#2#3{
  \left\{\begin{array}{ll}
    #2 & \mbox{if $#1$} \\
    #3 & \mbox{otherwise}
  \end{array}\right.}

\excludeversion{HIDE}
\iffull
\newenvironment{FULL}{\ifdraft\color{dkblue}\fi}{\Black}
\excludeversion{SHORT}
\includeversion{LONG}
\includeversion{FULL}
\else
\includeversion{SHORT}
\excludeversion{LONG}
\excludeversion{FULL}
\fi

\ifdraft

\newcommand{\displaycomment}[1]{\marginpar{\raggedright\scriptsize{#1}}}
\includeversion{DRAFT}
\else
\newcommand{\displaycomment}[1]{}
\excludeversion{DRAFT}
\fi

\newcommand{\fun}{Fun\xspace}

\newcommand{\target}{$\!{\int}\!$un\xspace}

\newcommand{\fsharp}{F\#\xspace}

\def\lstCaml{\lstset{language=[Objective]Caml,
  morekeywords=[1]{type,val,fun,let,in,ref,of,try,if,then,else,match,with,do,open,module,member,abstract,rec,assume,assert,init,sample,random,marginal,observe,weight,nil,yield,unit,bool,int,real,double,local,fail,fst,snd,case,inl,inr,elsif,blueC},
  morekeywords=[2]{public,interface,class},
  morekeywords=[3]{},%
  morekeywords=[4]{bn,fn,fv,dom,env,clauses},%
  morestring=[b]",
  sensitive=true,%
  columns=[l]fullflexible,
  texcl=true,
  mathescape=true,
  identifierstyle={\sffamily\small\maybecolor{dkgreen}},
  keywordstyle=[1]{\bfseries\maybecolor{dkblue}},
  keywordstyle=[2]{\bfseries\maybecolor{dkblue}},
  keywordstyle=[3]{\bfseries\maybecolor{black}},
  keywordstyle=[4]{\rmfamily\itshape},
  morecomment=[s]{(*}{*)},
  morecomment=[is]{(*---\ }{*)},
  morecomment=*[l][identifierstyle]{//},
  morecomment=*[l][identifierstyle]{///},
  rangeprefix=(*---\ ,
  includerangemarker=false,
  stringstyle=\ttfamily,
  commentstyle=\rmfamily\itshape,
  showspaces=false,
  showstringspaces=false,
  literate={/\\}{$\wedge\,$}{2} {\\/}{$\vee\,$}{2}
  {>>>}{$\Athen[]\,$}{1}
  {|->}{$\mapsto$}{1}
  {Bluec}{{\maybecolor{dkblue}c}}{1}
  {->}{$\rightarrow\,$}{1} {=>}{$\Rightarrow\,$}{1}
  {<-}{$\leftarrow\,$}{1}
  {<=>}{$\Leftrightarrow\,$}{3} {'a}{{\small$\alpha\,$}}{1}
  {'b}{{\small$\beta\,$}}{1} {'c}{{\small$\gamma\,$}}{1}
  {'d}{{\small$\delta\,$}}{1} {'e}{{\small$\epsilon\,$}}{1}
  {'f}{{\small$\phi\,$}}{1} {=def}{$\deq\ $}{3}
  {Fun.}{}{1}
  {GaussianFromMeanAndPrecision}{Gaussian}{1}
  {GammaFromShapeAndScale}{Gamma}{1}
  {~}{$\neg$}{1},
  tabsize=2,
  breaklines=true}}

\lstCaml
\let\ls\lstinline
\newcommand{\kw}[1]{\mbox{\normalfont\lstinline|#1|}}

\def\lstsnippet#1#2#3{\lstinputlisting[linerange=#2-#3]{Snippets/#1}}
\def\lstfrag#1/#2.{\lstsnippet{#1}{#2Begin}{#2End}}
\def\lstlearner#1/#2.{\lstinputlisting[linerange=#2Begin-#2End]{./Popl-#1}}

\newcommand{\GAP}{1ex}
\makeatletter
\lstnewenvironment{lstNumbered}[1][]{%
  \lstset{%
    #1,
    language=Caml,
    numbers=left,
    firstnumber=auto,
  }%
}{%
}
\makeatother

\makeatletter
  \newcommand{\addToLabel}[1]{%
    \protected@edef\@currentlabel{\@currentlabel#1}%
  }
\makeatother

\newtheorem{theoremI}{thm}
\newtheorem{propositionI}{prop}
\newtheorem{lemmaI}[propositionI]{lem}

\ifieee

\newenvironment{lem}[1][]{%
  \lemmaI[#1]%
  \ifx\relax#1\relax\else
    \addToLabel{ (#1)}
  \fi
}{%
  \endlemmaI
}

\newenvironment{thm}[1][]{%
  \theoremI[#1]%
  \ifx\relax#1\relax\else
    \addToLabel{ (#1)}
  \fi
}{%
  \endtheoremI
}

\fi

\ifieee

\else\ifjournal

\else

\fi\fi
{\begin{trivlist}\item[]{\normalsize\bf Restatement of #1}\hspace*{4mm}\it}%
  {\end{trivlist}}

\newcommand{\hbra}{
  \hbox to \columnwidth{\vrule width0.3mm height 1.8mm depth-0.3mm
    \leaders\hrule height1.8mm depth-1.5mm\hfill
    \vrule width0.3mm height 1.8mm depth-0.3mm}}
\newcommand{\hket}{
  \hbox to \columnwidth{\vrule width0.3mm height1.5mm
    \leaders\hrule height0.3mm\hfill
    \vrule width0.3mm height1.5mm}}

\newcommand{\ratio}{.35}

\newenvironment{display}[2][\ratio]{
  \begin{tabbing}
    \hspace{0.1em} \= \hspace{1.5em} \= \hspace{#1\linewidth-3.2em} \= \hspace{1.5em} \= \kill
    \textbf{#2}\\[-.8ex]
    \hbra\\[-.8ex]
  }
  {\\[-.8ex]\hket
  \end{tabbing}}

\newcommand{\entry}[2]{\>\>$#1$\>\>#2}
\newcommand{\clause}[3][]{\>$#2$\>#1\>#3}
\newcommand{\Category}[2]{\clause{#1::=}{#2}}
\newcommand{\smallCategory}[3]{\clause{#1::={#2}}{#3}}

\newcommand{\subclause}[1]{\>\>$#1$}

 \lstnewenvironment{displaylisting}[1]
   {~\\\noindent\textbf{#1}\\[-.8ex]\hbra\vspace{-2ex}}
   {\vspace{-2ex}\hket\vspace{1ex}}

\newcounter{rule}

\newcommand{\staterule}[4][]{%
  \refstepcounter{rule}%
  \addToLabel{(\textsc{#2})}\label{#2}%
  $\begin{array}[b]{@{}l@{}}%
    \mbox{(\textsc{#2})#1}\\%
    \begin{array}{@{}c@{}}
      #3\\
      \hline      \raisebox{0ex}[2.5ex]{\strut}#4%
    \end{array}
  \end{array}$}

\newcommand{\Set}[1]{\{#1\}}                    

\newcommand{\deq}{\mathrel{\smash[t]{\triangleq}}}

\renewcommand{\implies}{\Rightarrow}

\newcommand{\ol}[1]{\overline{#1}}

\newcommand{\qq}[1]{[\hspace{-.25ex}[#1]\hspace{-.25ex}]}

\newcommand{\ty}{:}

\renewcommand{\emptyset}{\varnothing}

\newcommand{\fv}{\operatorname{fv}}               
\newcommand{\dom}{\operatorname{dom}}             

\newcommand{\range}{\operatorname{range}}         

\newcommand{\emptyEnv}{\varepsilon}

\newcommand{\translatorFont}[1]{\mathcal{#1}}

\newcommand{\funF}[1]{\kw{#1}}

\newcommand{\typeF}[1]{\funF{#1}}

\newcommand{\metaF}[1]{\mathtt{#1}}

\newcommand{\iqq}[2][\;]{\translatorFont{I}\qq{#2}#1}

\newcommand{\realT}{\typeF{real}}
\newcommand{\boolT}{\typeF{bool}}
\newcommand{\intT}{\typeF{int}}
\newcommand{\unitT}{\typeF{unit}}

\newcommand{\PDist}[1]{#1}

\newcommand{\Vals}[2][\!]{\mathbf{V}_{#1 #2}}

\newcommand{\Int}{\mathbb{Z}}

\newcommand{\Real}{\mathbb{R}}

\newcommand{\typeOf}[1]{\operatorname{ty}(#1)}

\newcommand{\measure}[1]{\mathbb{P}_{#1}}

\newcommand{\D}{\mbox{\textnormal{Dist}}}

\newcommand{\Lright}[1][{}]{\kw{inr}\ #1}
\newcommand{\Lleft}[1][{}]{\kw{inl}\ #1}
\newcommand{\Llet}[3]{\kw{let}\ {#1}={#2}\ \kw{in}\ {#3}}
\newcommand{\Lmatch}[5]{\kw{match}\ {#1}\ \kw{with}\ \Lleft {#2}\to\ {#3} \mid \Lright {#4}\to\ {#5}}

\newcommand{\Lif}[3]{\kw{if}\ {#1}\ \kw{then}\ {#2}\ \kw{else}\ {#3}}

\newcommand{\isDet}[1]{#1\ \operatorname{pure}}

\newcommand{\lam}[2]{\lambda #1. \ #2} 

\newcommand{\Pqq}[2][\ \sigma]{\translatorFont{P}\qq{#2}{#1}}

\newcommand{\Lfst}[1]{\kw{fst}\ #1}
\newcommand{\Lsnd}[1]{\kw{snd}\ #1}
\newcommand{\Lexp}{\kw{exp}}
\newcommand{\Llog}{\kw{log}}

\newcommand{\Op}{\operatorname{op}}
\newcommand{\Det}{\operatorname{detOp}}

\newcommand{\Pleft}{\metaF{inl}}
\newcommand{\Pright}{\metaF{inr}}

\newcommand{\Peither}[1][\;]{\metaF{either}{#1}}

\newcommand{\Pmatch}[4][x]{\metaF{match}\ {#2}\ \metaF{with}\ \Pleft\ {#1}: {#3}\ \mid \Pright\ {#1}: {#4}}
\newcommand{\Plet}[3]{\metaF{let}\ {#1} = {#2}\ \metaF{in}\ {#3}}

\newcommand{\Preturn}[1][\;]{\metaF{return}{#1}}
\newcommand{\Pbind}[1][\;]{\ensuremath{{#1}{\metaF{>\!\!\!>\!=}}{#1}}}

\newcommand{\Athen}[1][\;]{\ensuremath{{#1}{\metaF{>\!\!\!>\!\!\!>}}{#1}}}

\newcommand{\Omit}[1]{}

\newcommand{\hasType}[3][\Gamma]{#1\vdash #2 : #3} 
\newcommand{\iverson}[1]{[#1]} 

\newcommand{\abs}[1]{\lvert{#1}\rvert}

\newcommand{\CASE}[1]{\STATE \textbf{case} #1\textbf{:} \begin{ALC@g}}
\newcommand{\ENDCASE}{\end{ALC@g}}

\newcommand{\DEFAULT}{\STATE \textbf{default:} \begin{ALC@g}}
\newcommand{\ENDDEFAULT}{\end{ALC@g}}
\newcommand{\DEFAULTLINE}[1]{\STATE \textbf{default:} }

\newcommand{\MCMC}{\textsc{mcmc}\xspace}

\begin{document}
\title[Deriving Probability Density Functions]{Deriving Probability Density Functions\\from Probabilistic Functional Programs}
\author[S.~Bhat]{Sooraj Bhat\rsuper a}
\address{{\lsuper a}Georgia Institute of Technology}

\author[J.\ Borgstr{\"o}m]{Johannes Borgstr{\"o}m\rsuper b}
\address{{\lsuper b}Uppsala University}
\email{johannes.borgstrom@it.uu.se}

\author[A.~D.~Gordon]{Andrew D. Gordon\rsuper c}
\address{{\lsuper c}Microsoft Research \and University of Edinburgh}
\email{adg@microsoft.com}

\author[C.~Russo]{Claudio Russo\rsuper d}
\address{{\lsuper d}Microsoft Research}
\email{crusso@microsoft.com}


\begin{abstract}
The \emph{probability density function} of a probability distribution is a fundamental 
concept in probability theory and a key ingredient in various widely used machine 
learning methods.  
However, the necessary framework for compiling probabilistic functional programs to 
density functions has only recently been developed.
In this work, we present a density compiler 
for a probabilistic language with failure and both discrete and continuous distributions,
and provide a proof of its soundness.
The compiler greatly reduces the development effort of domain experts, 
which we demonstrate by solving inference problems from various scientific applications, 
such as modelling the global carbon cycle, using a standard Markov chain Monte Carlo framework.
\end{abstract}
\maketitle

\newcommand{\pdf}{\textsc{pdf}\xspace}
\newcommand{\eg}{\emph{e.g.}\xspace}
\newcommand{\ie}{\emph{i.e.}\xspace}

\section{Introduction}

\emph{Probabilistic programming} promises to arm data scientists with
declarative languages for specifying their probabilistic models, while
leaving the details of how to translate those models to efficient
sampling or inference algorithms to a compiler.
Many widely used machine learning techniques that might be employed
by such a compiler use the \emph{probability density function} (\pdf) of
the model as input.
Such techniques include \emph{maximum likelihood} or \emph{maximum a
posteriori estimation}, \emph{$L2$ estimation}, \emph{importance
sampling}, and \emph{Markov chain Monte Carlo} (\MCMC) methods~\citep{scott2001parametric,bishop:book}.

However, despite their utility, density functions have been largely absent from
the literature on probabilistic functional programming~\citep{DBLP:conf/popl/RamseyP02,DBLP:conf/uai/GoodmanMRBT08,KS09:EmbeddedProbabilisticProgramming}.
This is because the relationship between programs and their density
functions is not straightforward: for a given program, the \pdf may not 
exist or may be non-trivial to calculate.
Such programs are not merely infrequent pathological curiosities but in fact
arise in many ordinary scenarios.
In this paper, we define, prove correct, 
and implement an algorithm for automatically computing {\pdf}s
for a large class of programs written in a rich probabilistic programming language.
An abridged version of this paper was published as~\citep{tacas13:densities}.
\paragraph{\textbf{Probability density functions.}} 

In this work, probabilistic programs correspond directly to {\em
probability distributions}, which are important because they are a
powerful formalism for data analysis.  However, many techniques we
would like to use require the {\em probability density function} of a
distribution instead of the distribution itself.  Unfortunately, not
every distribution has a density function.

{\bf\em Distributions.} One interpretation of a probabilistic program is that it is a simulation
that can be run to generate a random sample from some set
$\Omega$ of possible outcomes.  The corresponding probability distribution $\mathbb{P}$ characterizes the program
by assigning probabilities to different subsets of $\Omega$ ({\em events}).
The probability $\mathbb{P}(A)$ for a subset $A$ of $\Omega$
corresponds to the proportion of runs that generate an outcome in $A$,
in the limit of an infinite number of repeated runs of the simulation.

Consider for example a simple \emph{mixture of Gaussians}, here written 
in \fun \citep{fun-esop11},
a probabilistic functional language embedded within F\# \citep{SGC07:ExpertFSharp}.
\begin{lstlisting}
 if flip 0.7 then random(Gaussian(0.0, 1.0)) else random(Gaussian(4.0, 1.0))
\end{lstlisting}
The program above specifies a distribution on the real line ($\Omega$ is
$\mathbb{R}$) and corresponds to a generative process that flips a
biased coin and then generates a number from one of two Gaussian
distributions, both with standard deviation 1.0 but with mean either
0.0 or 4.0 depending on the result of the coin toss.  In this example,
we will be more likely to produce a value near 0.0 than near 4.0 because of the bias.
%
%
The probability $\mathbb{P}(A)$ for $A=[0,1]$, for instance, is the
proportion of runs that generate a number between $0$ and $1$.

{\bf\em Densities.}  A distribution $\mathbb{P}$ is a function that
takes subsets of $\Omega$ as input, but for many purposes it turns out
to be more convenient if we can find a function $f$ that takes
elements of $\Omega$ {\em directly}, while still somehow capturing the
same information as $\mathbb{P}$.

When $\Omega$ is the real line, we are interested in a function $f$ that satisfies
$\mathbb{P}(A) = \int_A f(x)\ dx$ for all intervals $A$, and we call $f$ the
\emph{probability density function} (\pdf) of the distribution $\mathbb{P}$.
In other words, $f$ is a function where the area under its curve on an
interval $A$ gives the probability $\mathbb{P}(A)$ of generating an outcome falling in
that interval.  The \pdf of our example (pictured below)
is given by the function
\[ f(x)=0.7 \cdot \kw{pdf_Gaussian(0.0, 1.0, x)} + 0.3 \cdot \kw{pdf_Gaussian(4.0, 1.0, x)} \]
where \lstinline{pdf_Gaussian(mean, sdev, $\cdot$)} is the \pdf of a Gaussian distribution with mean \lstinline{mean} and standard deviation \lstinline{sdev} (the famous ``bell curve'' from statistics).
\begin{center}
\includegraphics[width=0.95\columnwidth]{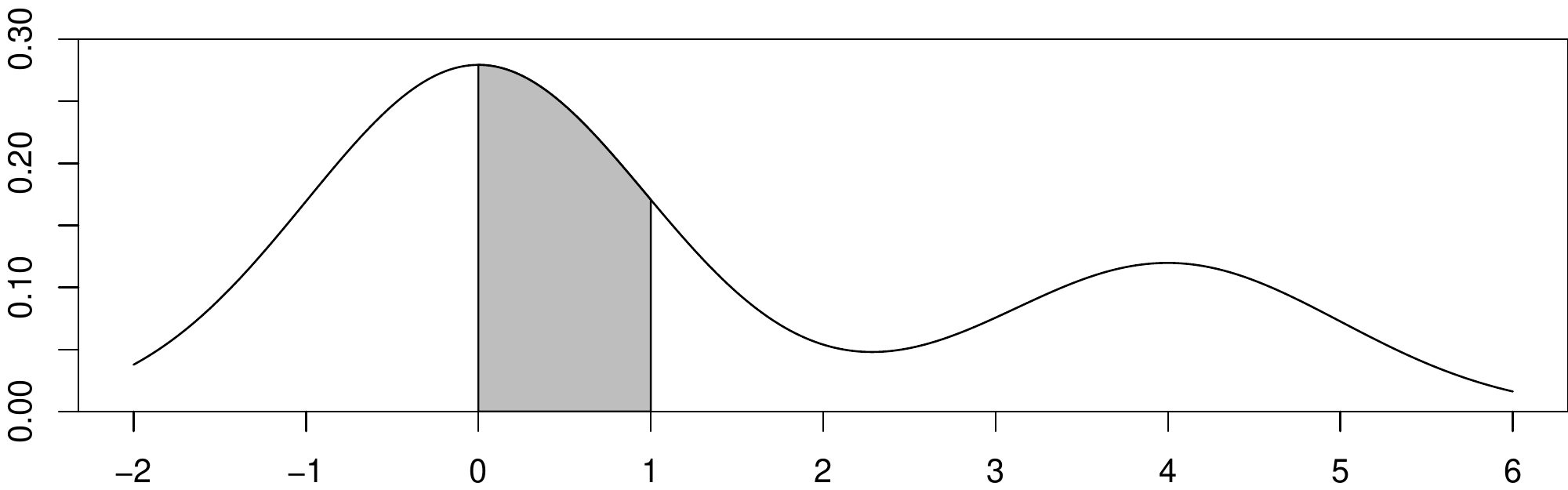}
\end{center}
The \pdf takes higher values where the generative process described
above is more likely to generate an outcome.  Now we see that the
aforementioned probability $\mathbb{P}(A)$ for $A=[0,1]$ is simply the
area under this curve between $0$ and $1$.  Note that, while there are
some loose similarities, the expression for the \pdf is different from
the expression comprising the source program.  In more complicated
programs, the correspondence with the \pdf is even less obvious.

{\bf\em Non-existence.} Sometimes a distribution does not have a \pdf.
For example, if we change the else-clause in our example to
return $4.0$ directly, instead of drawing from a Gaussian
with mean $4.0$, we get the following probabilistic
program, which does not have a \pdf:
\begin{lstlisting}
  if flip 0.7 then random(Gaussian(0.0, 1.0)) else 4.0
\end{lstlisting}
In short, the problem is that there is a non-zero amount of
probability mass located on a zero-width interval (the process now
returns $4.0$ with probability $0.3$), but integrals on such intervals
yield zero, so we would never find a function that could satisfy the
properties of being a \pdf.

It is not always obvious which program modifications ruin the property
of having a \pdf, especially for multivariate distributions (thus far
we have only given univariate distributions as examples).  This can be
a problem if one is innocently exploring different variations of the
model.
Details and examples are given by \cite{DBLP:conf/popl/BhatAVG12}, 
who provide the theory for addressing this problem, 
which we extend and implement in this work.

\paragraph{\textbf{The task of data analysis.}}

So far we have detailed {\em what} \pdf{}s are, but not {\em why} we
want them.  We motivate the desire by explaining one popular use-case
of \pdf{}s that arises when applying Bayesian learning to data
analysis.

In the previous examples, the programs specify fully-known
probability distributions. While there is indeed randomness in the
probabilistic behavior of the samples they generate, the nature of
this uncertainty is entirely known---we know the means and variances
of the Gaussians, as well as the bias between the two, thus we can
characterize the random behavior.

In real-world analysis tasks, we rarely have this luxury.  Instead, we
are often in the position of trying to figure out what the parameters
should be, given the data we see.
Thus, applications typically deal with {\em parameterized} models
(families of distributions indexed by a parameter), and they try to learn
something about which distributions in that family best explain the
observed data.  For example, in the following, \lstinline{moG} is a
parameterized model, indexed by the parameter \lstinline{(mA,mB)}:
\begin{lstlisting}
  let moG (mA,mB) = 
    if flip 0.7 then random(Gaussian(mA, 1.0)) else random(Gaussian(mB, 1.0))
\end{lstlisting}
It specifies an infinite number of distributions, one for each choice
of \lstinline{mA} and \kw{mB}.  

Now, as a data analyst, we may be presented a dataset that is a sequence
of numeric values, and we may also have some domain-specific reason to
believe that it can be well modelled as a biased mixture of Gaussians
as specified by \lstinline{moG}\footnote{Whether this is actually an
  appropriate modelling choice depends on whether it captures enough
  of the essence of the true, unknown data-generating process (\ie
  Nature), and is an entirely separate discussion.}.  We now face the
task of figuring out which choices of \lstinline{mA} and
\lstinline{mB} are likely.  Intuitively, if we see a clump of
datapoints around 0.0, we might be inclined to believe that the mean
of one of the Gaussians is 0.0.  Note that this is a one-dimensional,
probabilistic version of the problem of {\em clustering}.  The means
are the cluster centroids.

\paragraph{\textbf{Bayesian learning.}}

Bayesian modelling formalizes this task by requiring the modeller to
provide as input a {\em prior distribution} over the parameters and a
{\em generating distribution} over the data. These are used to
construct a {\em posterior distribution} over the parameters as the
output.

\clearpage
The prior distribution is a distribution over the possible values of
the parameters and captures our belief about what values they are
likely to take, before having seen any data (to a Bayesian, ``belief''
is synonymous with ``probability distribution'').  The following is
one possible prior:
\begin{lstlisting}
  let prior () = 
    let mA = random(Uniform(-10.0, 10.0)) in
    let mB = random(Uniform(-10.0, 10.0)) in
    (mA, mB)
\end{lstlisting}
This specifies that we are certain that the means lie between -10.0 and
10.0, but are otherwise uncertain, and our uncertainty about each
mean is uniformly distributed between -10.0 and 10.0.  This is of
course a very particular assertion and is hopefully informed by domain
knowledge.  In practice, in the absense of domain knowledge, we can
select a prior that reflects our open-mindedness about which values
the means can take, such as a pair of Gaussians with a high standard
deviation.  The prior {\em produces} a distribution over pairs of means as output,
rather than taking a pair of means as an input, as \lstinline{moG} does.

The generating distribution is a model of how we believe Nature is
generating a dataset given a specific choice of the parameter.  In our
example this is given by \lstinline{moG} together with
\begin{lstlisting}
  let gen n (mA,mB) = [| for i in 1 .. n ->  moG (mA,mB) |]
\end{lstlisting}
This specifies a (parameterized) model for a dataset that is generated
as an array of \kw{n} independent and identically distributed (i.i.d.)
values generated by \lstinline{moG}.

The posterior distribution is a distribution over the possible values
of the parameters and captures our belief about what values they are
likely to take, {\em after} seeing the data.  The posterior is related
to the prior and generating distributions by Bayes' rule, which
gives us a way to describe how the prior is updated with the observed
data to yield the posterior.  Intuitively, this update represents
the fact that our understanding about the world (our belief about the
parameters) evolves based on the data we see.  Bishop provides an
excellent account of Bayesian learning~\citep{bishop:book}.

\paragraph{\textbf{Use-case of \pdf{}s: Bayesian inference with \MCMC}} Unfortunately, while prior distributions and generating distributions
are often straightforward to work with (we have control over them as
the modeller), the posterior distributions often end up intractable or
unwieldy to work with (Bayes' rule dictates their form).

\emph{Markov chain Monte Carlo} (\MCMC) methods are one class of
techniques that let us actually do something productive with the
posterior distribution---\MCMC can be used to generate samples from
the posterior distribution.
The idea of \MCMC is to construct a Markov chain in the parameter space of the model, 
whose equilibrium distribution is the posterior distribution over model parameters. 
\cite{neal93} gives an excellent review of \MCMC methods. 
Here we use Filzbach \citep{Filzbach}, 
an adaptive \MCMC sampler based on the Metropolis-Hastings algorithm. 
All that is required for such algorithms is the ability to calculate
the posterior density given a set of parameters, up to proportion. 
The posterior does not need to be from a mathematically convenient
family of distributions.
Samples from the posterior can then serve as its representation,
or be used to calculate marginal distributions of parameters 
or other integrals under the posterior distribution.

\clearpage
The posterior density is a function of the \pdf{s} of the various
pieces of the model, so to perform inference using \MCMC, 
we also need functions to compute the \pdf{s}.
Below, \ls|pdf_moG| gives the \pdf of a single data point, 
while \ls|pdf_gen| gives the \pdf of an array of independent data points drawn from the same distribution (iid). 
\begin{lstlisting}
let pdf_prior (mA,mB) = pdf_Uniform(-10.0, 10.0, mA) * pdf_Uniform(-10.0, 10.0, mB)
let pdf_moG (mA,mB) x = 0.7 * pdf_Gaussian(mA, 1.0, x) + 0.3 * pdf_Gaussian(mB, 1.0, x)
let pdf_gen (mA,mB) xs = product [| for x in xs -> pdf_moG (mA,mB) x |]
\end{lstlisting}
(The \kw{product} function multiplies together the elements of an array, returning $1.0$ on the empty array.)
Filzbach and other \MCMC libraries require users to write these three functions\footnote{The actual implementation works with log-densities, as discussed in Section~\ref{sec:evaluation}.},
in addition to the generative probabilistic functions \kw{prior} and \kw{gen} 
(which are used for model validation).

The goal of this paper is to instead compile these density functions from the generative code. 
This relieves domain experts from having to write the density code in the first place, 
as well as from the error-prone task of manually keeping their model code and 
their density code in synch. 
Instead, both the \pdf and synthetic data are derived from the same declarative specification of the model.

\paragraph{\textbf{Contributions of this paper.}}
This work defines and applies automated techniques for computing densities
to real inference problems from various scientific applications.
The primary technical contribution is a \emph{density compiler} that is correct, useful, and relatively simple and efficient.  Specifically:
\begin{itemize}
\item 
We provide the first implementation of a density compiler based on the specification by~\cite{DBLP:conf/popl/BhatAVG12}.
We compile programs in the probabilistic language \fun (described in Section~\ref{sec:fun})
to their corresponding density functions (Section~\ref{sec:density-compiler}).
\item We prove that the compilation algorithm is sound (Theorem~\ref{thm:compiler-sound}).
This is the first such proof for any variant of this compiler.
\item We show that the compiler greatly reduces the development effort of domain experts 
by freeing them from writing tricky density code and that the produced code is
comparable in performance to density functions hand-coded by experts.  
Our evaluation is based on textbook examples and on models from ecology 
(Section~\ref{sec:evaluation}).
\end{itemize}
%

\section{Languages}
\label{sec:languages}
In order to describe the density compiler, we first specify its input
(source) and output (target) language.  Both languages are variants of
a simple first-order functional language where the results of subcomputations can be
bound to variables using a \kw{let} construct.
\subsection{\fun: Probabilistic Expressions (Review)}
\label{sec:fun}
\newcommand{\freshin}{\mathrel{\sharp}} 
\newcommand{\Pzero}{\metaF{zero}}
Our source language is a version of the core calculus \fun \citep{fun-esop11}, without observation. 
To mark certain program points as impossible, 
we add a \kw{fail} construct~\citep{KS09:EmbeddedProbabilisticProgramming}.
\fun is a first-order functional language without recursion that extends 
the language of~\cite{DBLP:conf/popl/RamseyP02},
and this version has a natural semantics in the sub-probability monad.
%
%
%
Our implementation efficiently supports a richer language with records and fixed-size arrays and array comprehensions, 
which can be given a semantics in this core (records and arrays can be encoded as tuples, 
and comprehensions of fixed size as their unrolling).

\subsubsection{Syntax and Types of \fun:}%
The language \fun has base types $\intT$, $\realT$ and $\unitT$, 
product types (denoting pairs), and sum types (denoting tagged unions).
A type is said to be \emph{discrete} if it does not contain $\realT$.
We let $c$ range over constant data of base type, $n$ over integers and $r$ over real numbers.
We write $\typeOf{c}=t$ to mean that constant $c$ has type $t$.

\begin{display}[.45]{Types of \fun:}
\smallCategory{t,u}{\kw{int} \mid \kw{real} \mid \kw{unit} \mid (t_1 * t_2) \mid (t_1 + t_2)}{}
\end{display}


We take $\boolT\deq\unitT+\unitT$, and let $*$ associate to the right.
We assume a collection of total deterministic functions on these types, 
including arithmetic and logical operators.
Each operation $\Op$ of arity $n$ has a signature of the form
\ls|val $\Op$: $t_1$ * $\cdots$ * $t_n$ -> $t_{n+1}$|. 
We also assume standard families of primitive probability distributions,
including the following. 
\begin{display}{Distributions:
    $\D: (t_1*\dots*t_n) \to \PDist{t}$}
\>$\kw{Bernoulli} : (\kw{real}) \to \PDist{\kw{bool}}$\\
\>$\kw{Poisson} : (\kw{real}) \to \PDist{\kw{int}}$\\
\>$\kw{Gaussian}: (\kw{real}*\kw{real}) \to \PDist{\kw{real}}$\\
\>$\kw{Beta}: (\kw{real}*\kw{real}) \to \PDist{\kw{real}}$\\
\>$\kw{Gamma}: (\kw{real}*\kw{real}) \to \PDist{\kw{real}}$
\end{display}
Above, the names $x_{i}$ of the arguments to the distributions are present for documentation only.
A $\kw{Bernoulli}(\textit{bias})$ distribution corresponds to a
coin flip with probability $\textit{bias}$ to come up $\kw{true}$.
The $\kw{Poisson}(rate)$ distribution describes the number of occurrences 
of independent events that occur at the given average \textit{rate}.
The \kw{Gaussian}$(\textit{mean}, \textit{stdev})$ distribution is
also known as the normal distribution; its \pdf has a symmetrical bell shape.
The \kw{Beta}$(a,b)$ distribution is a suitable prior 
for the parameter of \kw{Bernoulli} distributions, and intuitively
means that $a-1$
counts of \kw{true} and $b-1$ events of \kw{false} have been observed.
Similarly the \kw{Gamma}(\textit{shape},\textit{scale}) distribution is a suitable prior for 
the parameter of \kw{Poisson}. 
%
The parameters of distributions only make sense within certain ranges 
(e.g., the bias of the \kw{Bernoulli} distribution must be in the interval $[0,\,1]$). 
Outside these ranges, attempting to draw a value from the distribution (e.g., \kw{Bernoulli}$(2.0)$)
results in a failure (\kw{fail} below).
\begin{display}[.4]{Expressions of \fun:}
  \Category{V}{value}\\
  \entry{x}{variable}\\
  \entry{c}{scalar constant}\\
  \entry{(V,V)}{tuple constructor}\\
  \entry{\kw{inl}_u\; V}{left sum constructor}\\
  \entry{\kw{inr}_t\; V}{right sum constructor}\\[\GAP]
\Category{M,N}{expression}\\
\entry{x\mid c}{variable and scalar constant}\\
\entry{(M,N)\mid \kw{fst}\ M \mid \kw{snd}\ M} {pairing and projections from a pair}\\
\entry{\kw{inl}_u\; M \mid\kw{inr}_t\; M}{sum constructors}\\
\entry{
\kw{match}\ M\ \kw{with}\ \begin{prog}
\ \ \kw{inl}\ x_1 \to N_1
\mid \kw{inr}\ x_2 \to N_2
\end{prog}}{matching (scope of $x_i$ is $N_i$})\\
\entry{\Llet{x}{M}{N}}{let (scope of $x$ is $N$)}\\
\entry{\Op(M)}{primitive operation (deterministic)}\\
\entry{\kw{random}(\D(M))}{primitive distribution}\\
\entry{\kw{fail}_t }{failure}
\end{display}
The \kw{let} and \kw{match} statements bind their variables ($x,x_1,x_2$); 
we identify expressions up to alpha-renaming of bound variables.
Above, \kw{inl} (resp.~\kw{inr}) generates a value corresponding to the left (resp.~right) branch of a sum type.
Values of sum type are deconstructed by the \kw{match} construct, 
which behaves as either the left ($N_1$) or the right ($N_2$) branch depending on the result of $M$.

To ensure that a program has at most one type in a given typing environment, 
$\Lleft$ and $\Lright$ are annotated with a type (see \ref{Fun Inl} below). 
The expression \kw{fail} is annotated with the type it is used at.
These types are included only for the convenience of our technical development, and can usually be inferred given a typable source program: we omit these types where they are not used.
\kw{()} is the \kw{unit} constant.

A source language term $M$ is pure, written ``$\isDet{M}$'', iff $M$ does not contain any occurrence of \kw{random} or \kw{fail}.

We write 
\kw{Uniform} for \kw{Beta(1.0,1.0)}. 
In the binders of \kw{let} and \kw{match} expressions, 
we let~$\_$ stand for a variable that does not appear free in the scope of the binder.
We make use of standard sugar for \kw{let}, such as writing $M;N$ for $\kw{let}\ \_=M\ \kw{in}\ N$.
We write \kw{if} $M$ \kw{then} $N_1$ \kw{else} $N_2$ for $\Lmatch{M}{\_}{N_1}{\_}{N_2}$; this is most commonly used when $M$ is Boolean.
We let the tuple $(M_{1},M_{2},\dots,M_{n})$ stand for $(M_{1},(M_{2},\dots,M_{n}))$.
Similarly, we write $\Llet{x_{1},x_{2},\dots,x_{n}}VN$ for 
$\Llet{x_{1}}{\kw{fst}\;V}{\Llet{z}{\kw{snd}\;V}{\Llet{x_{2},\dots,x_{n}}zN}}$
when $z\freshin N$.

When $X$ is a term from some language (possibly with binders), 
we write $x_1,\dots,x_n\freshin X$ if none of the $x_i$ appear free in $X$.

We write $\Gamma \vdash M : t$ to mean that in the type environment
$\Gamma = x_1:t_1,\dots,x_n:t_n$ ($x_i$ distinct) the expression $M$ has type $t$.
Apart from the following, the typing rules are standard.
In \ref{Fun Inl}, \textsc{(Fun Inr)} (not shown) and \ref{Fun Fail}, 
type annotations are used in order to obtain a unique type.
In \ref{Fun Random}, a random variable drawn from a distribution of type 
$(x_1 \ty t_1*\dots*x_n \ty t_n) \to \PDist{t}$ has type $t$.

\begin{display}{Selected Typing Rules: $\Gamma \vdash M : t$}
\>
\staterule{Fun Inl}{
\Gamma \vdash M : t
}{
\Gamma \vdash \kw{inl}_u\ M : t+u
}
\qquad
\staterule{Fun Fail}{
}{
\Gamma \vdash \kw{fail}_t : t
}
\qquad
\staterule{Fun Random}{
\D: (t_1*\dots*t_n) \to \PDist{t}\\
\Gamma \vdash M \ty (t_1 * \dots * t_n)
}{
\Gamma \vdash \kw{random}(\D(M)) : t
}
\end{display}

Substitutions, ranged over by $\sigma,\rho$, 
are finite maps $[x_{1}\mapsto M_{1},\dots, x_{n}\mapsto M_{n}]$
from variables to pure expressions.  
We write $M\sigma$ for the result of substituting all free
occurrences of variables $x\in\dom(\sigma)$ in $M$ with $\sigma(x)$, avoiding
capture of bound variables.
To compose two substitutions with disjoint domains, 
we write $[x_{1}\mapsto M_{1},\dots, x_{n}\mapsto M_{n}]\sigma$ for
$[x_{1}\mapsto M_{1}\sigma,\dots, x_{n}\mapsto M_{n}\sigma]\cup\sigma$.
A substitution is called \emph{closed} if the expressions in its range do not contain any free variables.
A value substitution is a substitution where each expression in its range is a value.
Below, we define what it means for a closed value substitution to be a valuation for a type environment.
\begin{display}{Typing Rules for Closed Value Substitutions: $\Gamma\vdash\sigma$}
\>
\staterule{Subst Empty}
{
} {
\varepsilon\vdash[]
}
\qquad
\staterule{Subst Var}
{
\Gamma\vdash\sigma\qquad
\varepsilon\vdash V:t
} {
\Gamma,x\ty t\vdash\sigma[x\mapsto V]
}
\end{display}

\clearpage
There is a default value at each type $t$, written $0_{t}$,
that is returned from operations $\Op$ where they otherwise would be undefined, e.g.~$r/0.0=0_{\realT}=\kw{log}(-1)$.
\begin{display}{Default Value: $0_{t}$}
  \>
$0_{\unitT}:=()\qquad
0_{\intT}:=0\qquad
0_{\realT}:=0.0\qquad
0_{t*u}:=(0_t,0_u)\qquad
0_{t+u}:=\kw{inl}\;0_t$
\end{display}

\subsubsection{Semantics of \fun}\label{subsec-fun-semantics}
As usual, for precision concerning probabilities over uncountable sets, 
we turn to measure theory.
The interpretation of a type $t$ is the set $\Vals{t}$ of closed
values of type $t$ (real numbers, integers etc.).
Below we consider only Lebesgue-measurable sets of values, 
defined using the standard (Euclidian) metric, and ranged over by $A,B$.
Indeed, the power of the axiom of choice is needed to construct a non-measurable set~\citep{solovay70:lebesgue}.

A measure $\mu$ over $t$ is a function, from (measurable) subsets of $\Vals{t}$ to 
the non-negative real numbers extended with $\infty$, 
that is $\sigma$-additive, that is, $\mu(\emptyset)=0.0$ and $\mu(\cup_iA_i)=\Sigma_i\mu(A_i)$ 
if $A_1,A_2,\dots$ are pair-wise disjoint.
We write $\abs\mu$ for $\mu(\Vals{t})$;
the measure $\mu$ is called a probability measure if $\abs\mu=1.0$,
and a sub-probability measure if $\abs\mu\le1.0$.

We associate a default or \emph{stock} measure to each type, 
inductively defined as the counting measure on $\Int$ and $\{\kw{()}\}$, 
the Lebesgue measure on $\Real$, 
and the Lebesgue-completion of the product and disjoint sum, respectively,
of the two measures for $t*u$ and $t+u$.
In particular, the counting measure on a discrete type assigns measure $k$ to all sets of finite size $k$, 
and measure $\infty$ to all infinite sets.

If $f$ is a non-negative (measurable) function $t\to\realT$, 
we let $\int_{t} f$ be the Lebesgue integral of $f$ with respect to the stock measure on $t$
if the integral is defined, and otherwise 0.
This integral coincides with $\Sigma_{x\in\Vals{t}}f(x)$ for discrete types $t$, 
and with the standard Riemann integral (if it is defined) on $t=\realT$.
We write $\int_{t} f(x)\; dx$ for $\int_{t} \lambda x.f(x)$, 
and $\int_{t} f(x)\; d\mu(x)$ for Lebesgue integration with respect to the measure $\mu$ on $t$.
Below, we often elide the index $t$; indeed, we may consider any
function $t\to \realT$ as a function from the measurable space
$\uplus_{u}\Vals{u}$ that is zero except on $\Vals{t}$.

The Iverson brackets $[p]$ are 1.0 if predicate $p$ is true, and 0.0 otherwise.
We write $\int_Af$ for $\int \lambda x.[x\in A]\cdot f(x)$ when
$A\subset\Vals{t}$.
The function $g$ is a \emph{density} of $\mu$ (with respect to the stock measure) if
$\int_A 1\,d\mu(x)$ = $\int_A g$ for all $A$. 
If $\mu$ is a (sub-)probability measure, 
then we say that $g$ as above is its \pdf.

To turn expressions into density functions, we first need a way of interpreting a closed \fun expression $M$ as a sub-probability measure $\measure{M}$ over its return type.
Open \kw{fail}-free \fun expressions have a straightforward 
semantics~\citep{DBLP:conf/popl/RamseyP02} as computations in the probability monad~\citep{giry82.probMonad}.
In order to treat the \kw{fail} primitive, 
we use an existing extension \citep{modelLearner} of the semantics of \citet{DBLP:conf/popl/RamseyP02} to a richer monad: the sub-probability 
monad~\citep{DBLP:journals/entcs/Panangaden99}\footnote{Sub-probabilities 
are also used in our compilation of \kw{match} (and \kw{if}) statements, 
where the probability that we have entered a particular branch may be less than 1.}.
Compared to the operations of the probability monad, 
the sub-probability monad additionally admits a $\Pzero$ constant, yielding the zero measure.
To accommodate the zero measure, the carrier set
is extended from probability measures to sub-probability measures, i.e., admitting all $\mu$ with  $\abs\mu\le1$. 

Below we recapitulate the semantics of \fun by \citet{modelLearner}. Here $\sigma$ is a closed value substitution whose domain contains all the free variables of $M$, 
and $\Det(M)$ ranges over $\Op(M)$, 
$\kw{fst}\ M$,
$\kw{snd}\ M$,
$\Lleft[M]$ and
$\Lright[M]$.  We also let $\Peither f\; g\; (\kw{inl}\; V)\deq f \; V$ and $\Peither f\; g\; (\kw{inr}\; V) \deq g \; V$.
\clearpage

\begin{display}[.45]{Monadic Semantics of \fun with \kw{fail}: $\Pqq{M}$}
\clause{(\mu \Pbind f)\ A \deq \int\!f(x)(A)\,d\mu(x)}Sub-probability
monad's bind
\\
\clause{(\Preturn V)\ A \deq 1\text{~if~}V\in
  A\text{,~else~}0}Sub-probability monad's return
\\
\clause{\Pzero\ A \deq 0}Sub-probability monad's zero
\\[\GAP]\>
Below we assume that $z\freshin N,N_1,N_{2},\sigma$ and
$x,x_{1},x_{2}\freshin z,\sigma$.
\\[\GAP]
\clause{\Pqq{x}\deq \Preturn \sigma(x)}
\\\clause{\Pqq{c}\deq \Preturn c}
\\\clause{\Pqq{\Det(M)} \deq \Pqq{M}\Pbind\lambda x.\Preturn \Det(x)}
\\\clause{\Pqq{(M,N)} \deq \Pqq{M}\Pbind \lambda z.\Pqq{N}\Pbind\lambda w.\Preturn (z,w)}
\\\clause{\Pqq{\kw{let}\ x=M\ \kw{in}\ N} \deq 
  \Pqq{M} \Pbind \lambda z. \Pqq[{\ (\sigma[x\mapsto z])}]{N}}
\\[\GAP]
\clause{\Pqq{\Lmatch{M}{x_1}{N_1}{x_2}{N_2}}\deq  }
\\\clause{\qquad  \Pqq{M} \Pbind \Peither 
  (\lambda z. \Pqq[{\ (\sigma[x_1\mapsto z])}]{N_1})\;
  (\lambda z. \Pqq[{\ (\sigma[x_2\mapsto z])}]{N_2})
}
\\[\GAP]
\clause{\Pqq{\kw{random}(\D(M))} \deq \Pqq{M}\Pbind\lambda z.\mu_{\D(z)}}
\\\clause{\Pqq{\kw{fail}}\deq \Pzero}
\end{display}
We let the semantics of a closed expression $M$ be
$\measure{M}\deq\Pqq[\ \varepsilon]{M}$, where $\varepsilon$ denotes
the empty substitution.

\begin{lem}~
If $\Gamma \vdash M : t$ and $\Gamma\vdash\sigma$ then $\Pqq{M}$ is a sub-probability measure on type $t$.
\end{lem}
\proof
  By induction on $M$.
\qed

\subsection{Target Language for Density Computations}
For the target language of the density compiler, denoted \target, 
we use a pure version of \fun augmented with real-valued
first-order functions and stock integration.

\begin{display}[.48]{Expressions of the Target Language: $E,F$}
\Category{E,F}{target expression}\\
\entry{x\mid c}{variable and scalar constant}\\
\entry{(E,F)\mid \kw{fst}\ E \mid \kw{snd}\ E} {pairing and projections from a pair}\\
\entry{\kw{inl}_u\; E \mid\kw{inr}_t\; E}{sum constructors}\\
\entry{
\kw{match}\ E\ \kw{with}\ \begin{prog}
\ \ \kw{inl}\ x_1 \to F_1
\mid \kw{inr}\ x_2 \to F_2
\end{prog}}{matching (scope of $x_i$ is $F_i$})\\
\entry{\Llet{x}{E}{F}}{let (scope of $x$ is $F$)}\\
\entry{\Op(E)}{primitive operation}\\
\entry{\int_t \lam{(x_1,\dots,x_n)}E}{stock integration}
\end{display}
Above, the binders in \kw{let} and \kw{match} are as in Fun. Additionally,
in $\int_t \lam{(x_1,\dots,x_n)}E$ the variables $x_1, \dots,x_{n}$ bind into $E$.

If a Fun term $M$ is pure 
then $M$ is also an expression in the syntax of \target, 
and we silently treat it as such.
In particular, a Fun substitution $\sigma$ is also a valid \target substitution,
and substitution application $E\sigma$ for \target is defined in the same way as for Fun.

\clearpage
The typing rule involving integration is as follows.
The other typing rules are as in \fun.
\begin{display}{Typing Rule for Integration: $\Gamma \vdash E : t$}
\>
\staterule{Target Int}{
\Gamma,x_{1}:t_1,\dots,x_{n}:t_n \vdash E : \realT
}{
\Gamma \vdash \int_{t_1*\dots*t_n} \lam{(x_{1},\dots,x_{n})} E : \realT
}
\end{display}

\begin{lem}[Standard results for the type system of \target]
\label{lem:standard}$ $
\begin{enumerate}
\item \label{item:substitution} \emph{Substitution lemma}:
if $\Gamma,x:t,\Gamma'\vdash E:u$ and $\Gamma\vdash F:t$, then
$\Gamma,\Gamma'\vdash E[x\mapsto F]:u$.
\item \label{item:strengthening} \emph{Strengthening}: if $\Gamma,x:t,\Gamma'\vdash E:u$ and
  $x\freshin E$, then $\Gamma,\Gamma'\vdash E:u$.
\item \label{item:weakening} \emph{Weakening}: if $\Gamma,\Gamma'\vdash E:u$ and
  $x\freshin \Gamma,\Gamma'$, then $\Gamma,x:t,\Gamma'\vdash E:u$.
\end{enumerate}  
\end{lem}

\subsubsection{Semantics of \target}
\label{sec:targetsemantics}
The target language \target is equipped with a denotational semantics, 
written $\iqq{F} \sigma$ where $\sigma$ is a substitution of closed values for variables with $\dom(\sigma)\supseteq\fv(F)$.
We define this semantics by re-interpreting the monadic semantics of
Subsection~\ref{subsec-fun-semantics} with respect to the identity monad:
in this monad, $\Preturn[]$ is the identity function, and \texttt{bind} ordinary function application.
We rely on an auxiliary semantics $\iqq{\lam{(z_{1},\dots,z_{n})}E}\sigma$ that returns a function to be integrated.

\begin{display}[.45]{Identity Monad and Denotational Semantics of \target: $\iqq{\lam{(z_{1},\dots,z_{n})}E}\sigma$ and $\iqq{E}\sigma$}
\clause{V \Pbind f \deq f(V) }Identity monad's bind
\\
\clause{(\Preturn V) \deq V}Identity monad's return
\\[\GAP]{}
\clause{
  \iqq{\int_t \lam{(z_1,\dots,z_n)} F}\sigma \deq
  \condarray{ \mbox{the integral is well-defined} }{\int_t \iqq{\lam{(z_1,\dots,z_n)} F} \sigma}{0}
  } \\
(the other cases of $\iqq{E}\sigma$ are the same as the monadic semantics in Subsection~\ref{subsec-fun-semantics})\\[\GAP]
\clause{
  \iqq{\lam{(z_1,\dots,z_n)} F} \sigma \deq f}\\
\subclause{\text{where~}f(V_1,\dots,V_n) \deq \iqq F\sigma[z_1:=V_1,\dots,z_n:=V_n]\text{~and~}z_1,\dots,z_n\freshin \sigma}
\end{display}
We write $E\equiv F$ if there are $\Gamma,t$ such that
$\Gamma\vdash E:t$ and $\Gamma\vdash F:t$ and for all $\sigma$
such that $\Gamma\vdash \sigma$ we have $\iqq E\sigma=\iqq F \sigma$.\medskip

\begin{lem}~
\begin{enumerate}
\item If $\Gamma \vdash E : t$ and $\Gamma\vdash\sigma$ then $\iqq{E}\sigma$ is a value of type $t$.
\item If $\Gamma,x_{1}:t_1,\dots,x_{n}:t_n \vdash E : \realT$ and $\Gamma\vdash\sigma$ and $\dom(\sigma) \supseteq \fv(E)$
then $\iqq{\lam{(z_{1},\dots,z_{n})}E}\sigma$ is a function of type $t_1 * \dots * t_n \to \realT$.
\end{enumerate}
\end{lem}
\proof
(1) and (2) are proved jointly, by induction on $E$.
\qed

\section{The Density Compiler}
\label{sec:density-compiler}
\newcommand{\probCtx}{\Upsilon} 
\newcommand{\emptyCtx}{\epsilon}
\newcommand{\wfCtx}[2][\Gamma]{#1 \vdash #2\ \operatorname{wf}}
\newcommand{\newUnobs}[2]{#1 = #2}
\newcommand{\newObs}[3]{#1 = #2\ @\ #3}
\newcommand{\rands}[1][\probCtx]{\operatorname{rands}(#1)}
\newcommand{\consts}[1][\probCtx]{\sigma_{#1}}
\newcommand{\constEnv}[1][\probCtx]{{#1}\rvert_{=}}

\newcommand{\isConst}[2][\probCtx]{\isDet{#2} \quad \rands[#1]\freshin (#2\consts[#1])}
\newcommand{\isNotConst}[2][\probCtx]{\neg(\isDet{#2} \land \rands[#1]\freshin (#2\consts[#1]))}

\newcommand{\densityOf}[4][\probCtx]{#1; #3\vdash \operatorname{dens}(#2) \Rightarrow #4} 
\newcommand{\marginalOf}[4][\probCtx]{#1; #3\vdash \operatorname{marg}(#2) \Rightarrow #4} 

We compute the \pdf of a \fun program by compilation into \target.
Our compilation is based on that of~\cite{DBLP:conf/popl/BhatAVG12},
with modifications to treat \kw{fail} statements, 
\kw{match} (and general \kw{if}) statements,
pure (i.e., deterministic) \kw{let} bindings,  and integer arithmetic.


The compiler translates a well-typed \fun source term  $M$ 
into a function $\lam z F$ computing the density (PDF) of $M$.
Given an implementation of stock integration, the \target expression $F[z\mapsto V]$ may be
executed to evaluate the density of $M$ at any value $V$ of the type of $M$.
Like traditional compilers, our compiler is compositional and
deterministic, producing a unique translation if any at all (Lemma~\ref{lem:determinism}).
Unlike traditional compilers, our compiler is partial and will fail to produce a translation for some well-typed source terms.
In particular, if $M$ does not have a density function then the compiler will fail to produce an $F$. 
However, it may also fail if $M$ has a PDF, but the compiler is just not complete enough to compute it.
In particular, \kw{let}-bound expressions must either be pure or have a PDF, even if their result is not used.
The correctness statement for the compiler is given by Theorem~\ref{thm:compiler-sound}.

We will use a version of the \kw{moG} function from the introduction as a running example (Figure~\ref{fig:mog}), 
with some expansion in order to make use of more of the translation rules.
\begin{figure}[h]
  \centering
  \begin{minipage}[h]{0.5\linewidth}    
\begin{lstNumbered}
   let branch = random(Bernoulli(0.7)) in
   let temp = random(Gaussian(0.0, 1.0)) in
   match branch with 
     inl _ -> random(Gaussian(mA, 1.0)) 
    | inr _ -> 
       let result = temp + mB in 
       result
\end{lstNumbered}
    \end{minipage}
  \caption{Expanded model for a mixture of two Gaussians}
  \label{fig:mog}
\end{figure}

The structure of this section is as follows.
In Section~\ref{sec:outline} we provide an intuitive outline of the compilation.
We make preliminary definitions, such as the syntax of probability contexts $\Upsilon$, in Section~\ref{sec:contexts}.
We define the compiler itself in Section~\ref{sec:compilation-rules} in terms of a couple of judgments.
%
%
%
These judgments are inductively defined relations, but they in fact are partial functions and hence have a direct executable interpretation.
Finally, in Section~\ref{sec:compiler-correctness} we state and prove correctness of the compiler.
 
\subsection{Outline}
\label{sec:outline}
The simplest case in the density compilation is \kw{fail}, which
compiles to the function that always returns zero.
The compilation works on the \kw{let}-structure of the term:
a sequence of random \kw{let}s, as in
$\Llet{x_{1}}{\kw{random}(\D_{1}(V_{1}))}{\dots\kw{in}\;(x_1,\dots,x_{n})}$
is compiled to the product of the \pdf{s} of the distributions
$\D_{1},\dots,\D_{n}$, following the chain rule of probability.

If the sequence of \kw{let}s instead has a discrete deterministic return expression $M$,
then $M$ has a probability for each possible value $V$. 
This probability is computed by integrating the joint \pdf of $x_1,\dots,x_{n}$ 
over the set of values where $M$ evaluates to $V$.
A continuous deterministic return expression $M$ is treated as a
mathematical function $f_{M}$, using the change of variables rule of integration.
In the one-dimensional case, if $f_{M}(x)$ has inverse $f^{-1}$, 
the \pdf of $f_{M}$ at $r$ is given by the \pdf of $x$ at $f^{-1}(r)$,
multiplied with the derivative of $f^{-1}(r)$.
Another simple case is projection $M=x_{n}$, where we simply integrate the joint \pdf
over the set of all possible values for the other variables $x_1,\dots,x_{n-1}$.  

If a distribution $\D$ returns a sum type (e.g., \kw{Bernoulli}) we
can write $\D_{l}$ for the subdistribution yielding only the left part
of the sum, and $\D_{r}$ for the right part.
By additivity of probability, 
we can compile the \kw{match} expression 
$\Lmatch{\kw{random}(\D(V))}{y}{M}{z}{N}$ 
to the sum of the \pdf{s} of
$\Llet{y}{\kw{random}(\D_{l}(V))}M$ and $\Llet{z}{\kw{random}(\D_{r}(V))}N$.

In a nested \kw{let}, such as
$\Llet{x_{i}}{(\Llet{y_{1}}{\kw{random}(\D(V))}{\dots\ \kw{in}\;M_{i})}}\dots$, 
the expression bound to $x_{i}$ denotes some subprobability distribution.
We compute its \pdf by recursively compiling the
inner $\kw{let}$ sequence, holding $x_{1},\dots,x_{i-1}$ fixed.
Pure \kw{let}s, as in $\Llet{x}{M}{N}$ where $M$ is pure,
have the same \pdf as $N[x\mapsto M]$. The compilation algorithm applies the
substitution lazily to avoid introducing unnecessary copies of $M$.

\subsection{Probability contexts}
\label{sec:contexts}
The density compilation is based on the let-structure of the expression.
Variables that are bound in outer lets are referred to as parameters,
and are treated as constants.
A \emph{probability context} gathers the variables that are bound in the current sequence of lets,
together with the pure expressions defining the deterministic variables.

\begin{display}{Probability Context: $\probCtx$}
\Category{\probCtx}{probability context}\\
\entry{\emptyCtx}{empty context}\\
\entry{\probCtx, x}{random variable}\\
\entry{\probCtx, x=E}{deterministic variable}
\end{display}

\begin{exa} \label{ex:probctx}
  The probability context at line 7 of Figure~\ref{fig:mog} is 
  $\probCtx_{7}:={}$\kw{branch, temp, result}${}={}$\kw{temp + mB},
  containing two random variables and one deterministic variable.
\end{exa}

For a probability context to be well-formed, it has to be well-scoped and well-typed. 
\begin{display}{Well-formed probability context: $\wfCtx{\probCtx}$}
\>
\staterule{Env Empty}
{}
{\wfCtx{\emptyCtx}}

\qquad

\staterule{Env Var}
{\wfCtx{\probCtx}\qquad 
\hasType{x}{T}\qquad 
x\freshin\probCtx}
{\wfCtx{\probCtx, x}}

\qquad

\staterule{Env Const}
{
  \begin{array}{cc}
\wfCtx{\probCtx} &\qquad \hasType x T \\ 
x\freshin\probCtx&\qquad\hasType E T
\end{array}
}
{\wfCtx{\probCtx, {x}={E}}}
\end{display}
\begin{exa} \label{ex:probctx-wellformed}
  The probability context $\probCtx_{7}$ is well-formed: $\wfCtx[\Gamma_{7}]{\probCtx_{7}}$, where the type context\linebreak $\Gamma_{7}:=\kw{mA}{:}\realT,\kw{mB}{:}\realT,\kw{branch}{:}\boolT,\kw{temp}{:}\realT,\kw{result}{:}\realT$ also contains the types of the parameters \kw{mA} and \kw{mB}. 
\end{exa}

Given a well-formed context $\probCtx$, we can extract the random variables $\rands$,
and an idempotent substitution $\consts$ (i.e.,
$E\consts=(E\consts)\consts$ always) that gives values to the deterministic variables.
\begin{display}{Random variables $\rands$ and values of deterministic variables $\consts$
}
\>$
\begin{array}{rclrclrcl}
\rands[\emptyCtx] &\deq&\varepsilon
&\consts[\emptyCtx] &\deq&[]
\\[\GAP]
\rands[\probCtx, x] &\deq&\rands,x\quad
&\consts[\probCtx, x] &\deq&\consts
\\[\GAP]
\rands[\probCtx, x=E] &\deq&\rands
&\consts[\probCtx, x=E] &\deq&[x\mapsto E]{\consts}\quad
\end{array}$
\end{display}

\begin{exa} \label{ex:rands}
We have $\rands[\probCtx_{7}]=\kw{branch},\kw{temp}$ and  $\consts[\probCtx_{7}]=[\kw{result}\mapsto\kw{temp}+\kw{mB}]$.
\end{exa}

\begin{lem}
  If $\wfCtx{\probCtx}$ then $\dom(\consts)\freshin\range(\consts)$
\end{lem}
\proof
  By simultaneous induction on the derivations of $\wfCtx{\probCtx}$ and $\consts$.
\qed

\subsection{Compilation rules}
\label{sec:compilation-rules}
A probability context $\probCtx$ is used together with a density expression ($E$ below),
which is an open term that expresses the joint density of 
the random variables in the context 
and the constraints that have been collected when choosing branches in \kw{match} statements.
Intuitively, the density expression $E$ is the body of the \pdf of the current branch.
The main judgment of the compiler is $\densityOf{M}{E}{\lam z F}$, 
which associates a \fun term $M$ with its density function $z \mapsto F$.
Parameters may occur free in $F$, and $z$ binds into $F$.
The auxiliary judgment $\marginalOf{\{x_1,\dots,x_k\}}{E}F$ yields a density expression $F$ 
for the variables in its argument, 
marginalizing (i.e., integrating) out all other random variables in $\probCtx$ from $E$.
\begin{display}[.25]{Inductively Defined Judgments of the Compiler:}
\clause{ \densityOf{M}{E}{\lam z F} }{in $\probCtx;E$ the function $\lam z F$ gives the \pdf of $M$}\\
\clause{ \marginalOf{X}{E}F }{in $\probCtx;E$ expression $F$ gives the
density of the variables in $X$}
\end{display}

We begin the description of the compiler proper with the following judgment of marginal density,
which computes an expression for the joint marginal \pdf of the random variables in its argument. 
The variables in the argument are free in the computed expression.
Below, we write $x_{1},\dots,x_{n}\setminus Y$ for the tuple of variables
arising from $x_{1},\dots,x_{n}$ by deleting all instances of
variables in $Y$, and dually for $x_{1},\dots,x_{n}\cap Y$.
\begin{display}{Marginal Density: $\marginalOf{X}{E}F$}
\>\staterule{Marginal}
  {
   X\subseteq\rands\qquad \rands \setminus X = y_1,...,y_n 
  }
  {\marginalOf{X}E{\int\lam{(y_1,...,y_n)} E\consts}}
\end{display}
Here we first substitute in the deterministic \kw{let}-bound variables, as given by $\consts$,
and then integrate out the remaining random variables ${y_1,...,y_k}$.
In the definition of the compiler, marg($X$) is also used with $X=\emptyset$, 
to compute the probability of being in the current branch of the program.

\begin{exa} \label{ex:marginalof}
  Here $\marginalOf[\probCtx_{7}]{\{\kw{temp}\}}E{F_{\kw{temp}}}$ where \[F_{\kw{temp}}:={
  \int\lambda(\kw{branch}).E [\kw{result}\mapsto\kw{temp}+\kw{mB}]}\,,\]
  which will be used when computing the \pdf of the variable \kw{result} on line 7 (cf.~Examples~\ref{ex:basecases},~\ref{ex:translate}).
\end{exa}

The main judgment of the compiler is the dens judgment $\densityOf{M}{E}{\lam{z}F}$, 
which gives the density $z\mapsto F$ of $M$ in the current context $\probCtx$,
where $E$ is the accumulated body of the density function so far.
In this judgment, $z$ is binding into $F$.
We introduce fresh variables during the compilation:
in the rules below we assume that $z,w\freshin \probCtx,E,M$.
\begin{display}{Density Compiler, base cases: $\densityOf{M}{E}{\lam{z}F}\qquad\qquad\qquad$ }    
\>
\staterule{Var Det}
  {(x=E')\in\probCtx\quad \densityOf{E'}{E}{\lam{z}F}}
  {\densityOf{x}{E}{\lam{z}F}}

\qquad \qquad


\staterule{Var Rnd}
  {x\in\rands\quad \marginalOf{\{x\}}{E}{F}}
  {\densityOf{x}{E}{\lam{x}F}}

\\[\GAP]\>

\staterule{Constant}
  {\varepsilon\vdash V:t\qquad t\ \text{discrete}\qquad\marginalOf{\emptyset}EF}
  {\densityOf{V}{E}{\lam{z} F\cdot\iverson{z=V}}}

\qquad \quad

\staterule{Fail}
{}
{\densityOf{\kw{fail}}E{\lam{z}0.0}}

\end{display}
For a deterministic variable, \ref{Var Det} recurses into its definition.
The rule \ref{Var Rnd} computes the marginal density of a random variable using the marg judgment.
%
%
The \ref{Constant} rule states that the probability density of a discrete constant $V$
(built from sums and products of integers and units)
is the probability of being in the current branch at $V$, and 0 elsewhere.
Note the absence of a rule for real number constants, since they do not possess a density.
The \ref{Fail} rule gives that the density of \kw{fail} is zero.

\begin{exa} \label{ex:basecases}
  By \ref{Var Rnd} we get $\densityOf[\probCtx_{7}]{\kw{temp}}E{\lam{\kw{temp}}F_{\kw{temp}}}$ as computed in Example~\ref{ex:marginalof}.  

  To compute the \pdf at line 7, \ref{Var Det} yields that $\densityOf[\probCtx_{7}]{\kw{result}}E{\lam z F_{7}}$ 
  where $\densityOf[\probCtx_{7}]{\kw{temp} + \kw{mB}}E{\lam z F_{7}}$ is computed in Example~\ref{ex:translate}.
\end{exa}

\begin{display}{Density Compiler, \kw{random} variables : $\densityOf{M}{E}{\lam{z}F}$ }    
\>

\staterule{Random Const}
  {\isConst M \qquad\marginalOf{\emptyset}EF}
  {\densityOf{\kw{random}(\D(M))}{E}{\lam{z} F\cdot\textsc{pdf}_{\D(M\consts)}}(z)}

\\[\GAP]\>

\staterule{Random Rnd}
  {\neg(\isDet{M} \land \rands\freshin (M\consts))\qquad \densityOf{M}{E}\lam w F}
  {\densityOf{\kw{random}(\D(M))}{E}{\lambda z.\int\lambda w.F\cdot\textsc{pdf}_{\D(w)}(z)}}
\end{display}
In \ref{Random Const}, a random variable drawn from 
a primitive distribution with a constant argument 
has the expected \pdf
(multiplied with the probability that we are in the current branch).
Its precondition that $\isDet{M}$ and $\rands\freshin (M\consts)$ intuitively means that $M$ is constant under $\probCtx$.
\ref{Random Rnd} treats calls to \kw{random} with a random argument by
marginalizing over the argument to the distribution. 
We here require that each primitive distribution $\D$ has a \pdf for each value $w$ of its arguments, denoted $\textsc{pdf}_{\D(w)}$.
\begin{exa} \label{ex:random}
  Using rule \ref{Random Const}, we can compute the density at line 4 as\\
  $\densityOf{\kw{random}(\kw{Gaussian}(\kw{mA},1.0))}{E}{\lam{z} F_{\kw{then}}\cdot\textsc{pdf}_{\kw{Gaussian}(\kw{mA},1.0)}(z)}$
  where $F_{\kw{then}}$ intuitively yields the probability of being in the current branch, 
  and is computed using \ref{Marginal} as $\int\lambda(\kw{temp},\kw{branch}).E$.
\end{exa}

\begin{display}{Density Compiler, rules for tuples: $\densityOf{M}{E}{\lam z F}\qquad\qquad\qquad$ }    
\>

\staterule{Tuple Var}
  {\marginalOf{\{x_1,...,x_k\}}EF\qquad k\ge 2 \qquad x_1,...,x_k\text{~distinct}}
  {\densityOf{(x_1,...,x_k)}E\lam{z}\Llet{x_1,...,x_k}zF}

\\[\GAP]\>

\staterule{Tuple Proj L}
  {\densityOf{M}{E}\lam{w}F}
  {\densityOf{\Lfst{M}}E{\lam{z_1} \int \lam{z_2} \Llet{w}{(z_1,z_2)}F}}

\\[\GAP]\>

\staterule{Tuple Proj R} 
  {\densityOf{M}{E}\lam{w}F}
  {\densityOf{\Lsnd{M}}E{\lam{z_2} \int \lam{z_1} \Llet{w}{(z_1,z_2)}F}}
\end{display}
The rule \ref{Tuple Var} computes the joint marginal density of a tuple of variables\footnote{Joint marginal densities for tuples of expressions can be computed 
if those expressions are conditionally independent~\citep{DBLP:conf/popl/BhatAVG12}.
As an example, $(x, y+3)$ has a $\pdf$ whenever $(x,y)$ does.
However, the rules in this paper do not support such expressions, 
to avoid additional complexity.}.
The rules \ref{Tuple Proj L} and \ref{Tuple Proj R} integrate out the right or the left component of a pair, respectively.
\begin{display}{Density Compiler, \kw{let}: $\densityOf{\Llet{x}{M}{N}}{E}{F}$ }  
\>
\staterule{Let Det}
  {\isDet M \qquad\densityOf[\probCtx,x=M]{N}{E}\lam z F}
  {\densityOf{\Llet{x}{M}{N}}{E}\lam z F}

\qquad

\staterule{Let Rnd}
  {
    \begin{array}{cr}
      \neg(\isDet M) \quad&
    \densityOf[\emptyCtx]{M}{1}\lam x F_1\\
    &\densityOf[\probCtx,x]{N}{E\cdot F_1}{\lam z F_2}
  \end{array}
}
  {\densityOf{\Llet{x}{M}{N}}{E}{\lam z F_2}}

\end{display}
The rule \ref{Let Det} simply adds a pure let-binding 
to the context. 
%
In \ref{Let Rnd}, we compute the density of the let-bound variable 
in an empty context,
and multiply it into the current accumulated density 
when computing the density of the body.
\begin{exa} \label{ex:let}
  The density expression for the program fragment in Figure~\ref{fig:mog} is computed using \ref{Let Rnd} as 
  $F_{2}$ where $\densityOf[\kw{branch}]{N}{1\cdot F_{\kw{branch}}}{\lam z F_2}$, the expression $N$ is lines 2-7, and
  $F_{\kw{branch}}=(\int\lambda().1)\cdot\textsc{pdf}_{\kw{Bernoulli}(0.7)}(z)$ 
  is computed by
  $\densityOf[\emptyCtx]{\kw{random(Bernoulli(0.7))}}{1}\lam z F_{\kw{branch}}$
  using previously seen rules.
  Here
  $F_{\kw{branch}}\equiv \textsc{pdf}_{\kw{Bernoulli}(0.7)}$, since $\iqq{\int\lambda().1}\sigma=1$.

  The \kw{let} expression on line 2 of Figure~\ref{fig:mog} is also handled by \ref{Let Rnd}, 
  while the one on line 6 is handled by \ref{Let Det} since $\isDet{\kw{temp} + \kw{mB}}$.
\end{exa}

For deterministic matches we use four deterministic operations, 
  which we assume do not occur in source programs.
We let $\kw{isL}:t+u\to\realT$ be the indicator function for the left branch defined as $\kw{isL}(V):=\Pmatch V{1.0}{0.0}$, and dually for \kw{isR}.
To destruct a value of sum type we use 
$\kw{fromL}:t+u\to t$ defined as $\kw{fromL}(V):=\Pmatch{V}{x}{0_t}$, 
and its dual \kw{fromR}.

\begin{display}{Density Compiler, rules for sums and \kw{match}: $\densityOf{\kw{match}\ M\ \kw{with}\dots}{E}{\lam z F}$ }    
\>
\staterule{Sum Con L}
  {\densityOf{M}{E}\lam z F}
  {\densityOf{\Lleft{M}}{E}{\lam w \Lmatch{w}{z}{F}{\_}{0}}}

\\[\GAP]\>  

\iffull
\staterule{Sum Con R}
  {\densityOf{M}{E}\lam z F}
  {\densityOf{\Lright{M}}{E}{\lam w \Lmatch{w}{\_}{0}{z}{F}}}
\\[\GAP]\>  
\fi

\staterule{Match Det}
{
  \begin{array}{cc}
\isDet M &\qquad
  \densityOf[\probCtx,x_1=\kw{fromL}(M)]{N_1}{E\cdot\kw{isL}(M)}{\lam z F_1}\\
  &\qquad\densityOf[\probCtx,x_2=\kw{fromR}(M)]{N_2}{E\cdot\kw{isR}(M)}{\lam z F_2}
\end{array}}
{\densityOf{\Lmatch{M}{x_1}{N_1}{x_2}{N_2}}{E}
  {\lam{z} F_1 + F_2}
}

\\[\GAP]\>
\staterule{Match Rnd}
{ 
  \begin{array}{cc}
\neg(\isDet M) &\qquad
  \densityOf[\probCtx,x_1]{N_1}{E\cdot\Llet{w}{\kw{inl}\ x_1}F}{\lam z F_1}\\
  \densityOf[\emptyCtx]{M}{1}\lam w F&\qquad
  \densityOf[\probCtx,x_2]{N_2}{E\cdot\Llet{w}{\kw{inr}\ x_2}F}{\lam z F_2}
  \end{array}}
{\densityOf{\Lmatch{M}{x_1}{N_1}{x_2}{N_2}}{E}
  {\lam{z} F_1 + F_2}}

\\[\GAP]\>
\staterule{fromL}
{(x=\kw{fromL}(M))\in\probCtx\qquad\densityOf ME\lam w F}
{\densityOf{\kw{fromL}(M)}E{\lam z \Llet w{\kw{inl}\ z} F}}

\\[\GAP]\>
\staterule{fromR} 
{(x=\kw{fromR}(M))\in\probCtx\qquad\densityOf ME\lam w F}
{\densityOf{\kw{fromR}(M)}E{\lam z \Llet w{\kw{inr}\ z} F}}
\end{display}

\ref{Sum Con L} states that the density of $\kw{inl}\;M$ is the density of $M$ in the left branch of a sum, and 0 in the right.
Its dual is \ref{fromL}.
\ref{Match Det} is based on \ref{Let Det},
and we additionally multiply the constraint that we are in the correct branch 
($\kw{isL}(M)$ or $\kw{isR}(M)$) with the joint density expression.
We employ the functions \kw{fromL} and \kw{fromR} and their associated
rules \ref{fromL} and \ref{fromR}
to avoid additional calls to \ref{Match Det} arising from \ref{Var Det}
if the compilation of the density of $N_{i}$ requires computing the density of the match-bound variable $y_{i}$, as in $\Lmatch{\kw{fst}\ z}{y_1}{y_1}{y_2}{y_2}$.
%
Since we assume that \kw{fromL} and \kw{fromR} do not appear in source programs, 
these rules are only ever used in the case described above.
The \ref{Match Rnd} rule is based on \ref{Let Rnd}, 
and we again multiply in the constraint that we are in the left (or right) branch of the \kw{match}.
\begin{exa}\label{ex:match}
  The \kw{match} selector on line 3 is a pure expression, so
  rule \ref{Match Det} applies. For the left branch,
  we let $E_{4}\equiv \textsc{pdf}_{\kw{Bernoulli}(0.7)}(\kw{branch})\cdot\textsc{pdf}_{\kw{Gaussian}(0,1.0)}(\kw{temp}) \cdot\kw{isL}(\kw{branch})$ and
$\probCtx_{4} = \kw{branch},\kw{temp},\_=\kw{fromL}(\kw{branch})$ and  compute \[\densityOf[\probCtx_{4}]{\kw{random(Gaussian(mA, 1.0))}}{E_{4}}{\lam z F_4}\]
  where $F_{4}$ is computed using \ref{Random Const} and \ref{Marginal} as
  \[\left(\int\lambda(\kw{branch},\kw{temp}).E_{4}\right)\cdot\textsc{Pdf}_{\kw{Gaussian}(\kw{mA,1.0})}(z)\]
Here $\iqq{\int\lambda(\kw{branch},\kw{temp}).E_{4}}\sigma= 0.7$, so the contribution of the left branch to the \pdf of the \kw{match} is the \pdf of the branch scaled by the probability of entering the left branch. In general, this holds when the branch expression is independent from the body of the branch. 

For the right branch, see Example~\ref{ex:translate}. We then obtain the \pdf of the match as the sum of the {\pdf}s of the two branches.
\end{exa}

Our implementation of the compiler uses the following derived rules for \kw{if} statements
where the branching expression is of type $\kw{bool}$, and does not treat other sum types nor matches. 
\begin{display}{Derived rules for \kw{if} statements}
\>\staterule{If Det}
  {
  \begin{array}{c@{\qquad}c}
\isDet{M} &
    \densityOf{N_1}{E\cdot\kw{isL}(M)}{\lam z F_1}  \\ &
    \densityOf{N_2}{E\cdot\kw{isR}(M)}{\lam z F_2} 
\end{array}}
  {\densityOf{\Lif{M}{N_1}{N_2}}E
    {\lam{z} F_1 + F_2}}
  \iffull
    \\[\GAP]\>
\staterule{If Rnd}
  { 
  \begin{array}{c@{\qquad}c}
\neg(\isDet M)&
    \densityOf{N_1}{E\cdot\Llet w{\kw{true}}F}{\lam z F_1} \\
    \densityOf[\emptyCtx]{M}{1}\lam w F&
    \densityOf{N_2}{E\cdot\Llet w{\kw{false}}F}{\lam z F_2} 
\end{array}
}
  {\densityOf{\Lif{M}{N_1}{N_2}}E
    {\lam{z} F_1 + F_2}} \fi
\end{display}

\begin{exa} \label{ex:if} 
  Since the match-bound variables $\_$ (lines 4 and 5) do not appear in the bodies of the match branches,
  we can instead use rule \ref{If Det} to avoid adding them to the probability context when computing the \pdf of the body 
  (cf.~Example~\ref{ex:probctx}).  
\end{exa}

\begin{display}{Density compiler, discrete operations : $\densityOf{f(M)}{E}{F}$}
\>
\staterule{Discrete}
  { \isDet{M} \qquad\marginalOf{\{x_1,\dots,x_n\}}EF
    \qquad f\not\in\{\kw{fromL},\kw{fromR}\}
 \\f: t\to u \qquad
    u\text{~discrete} \qquad 
   \rands\cap\fv(M\consts) = x_1,\dots,x_n}
  {\densityOf{f(M)}E{}\lam{w} \int\lam{(x_1,\dots,x_n)} F\cdot[w=f(M\consts)] }
\end{display}
The \ref{Discrete} rule for discrete operations, 
such as logical and comparison operations 
and integer arithmetic, 
computes the expectation of an indicator function 
over the joint distribution of 
the random variables occurring in the expression.

For numeric operations on real numbers we mimic the change of variable rule of integration (often summarized as ``$dx =\frac{dx}{dy}dy$''), 
multiplying the density of the argument 
with the derivative of the inverse of the operation.  
For operations of more than one argument (e.g., \ref{Plus Rnd} below), we instead use the matrix
volume of the Jacobian matrix of the inverse operation~\citep{ben-israel99:matrixVolume}.
We only require that the operation is invertible on a restricted domain, 
namely where the \pdf of its argument is non-zero.
This is exemplified by the following rules.

\begin{display}{Density compiler, numeric operations on reals : $\densityOf{f(M)}{E}{F}$}
\>

\staterule{Inverse}
  { \densityOf{M}{E}\lam w F }
  {\densityOf{1/M}E{\lam{z} (\Llet{w}{1/z}F) \cdot (1/ z^2)}}

\\[\GAP]\>
\staterule{Exp}
  {\densityOf{M}{E}\lam w F}
  {\densityOf{\Lexp(M)}E{\lam{z} \kw{if}\;z> 0.0\ \kw{then} (\Llet{w}{\Llog(z)}F)\cdot(1/z)\ \kw{else}\ 0.0}}

\\[\GAP]\>
\staterule{Log}
 {\densityOf{M}{E}\lam{w}F
\qquad
\neg\exists \sigma, r<0,c\neq0. \iqq{\Llet w r F}\sigma = c}
 {\densityOf{\Llog(M)}E{\lam{z} (\Llet{w}{\Lexp(z)}F)\cdot\Lexp(z)}}

\\[\GAP]\>
\staterule{Scale}
 { c\neq 0\qquad \densityOf{M}{E}\lam w F }
 {\densityOf{c\cdot M}E{\lam{z} (\Llet{w}{z/c}F) \cdot (1/ \kw{abs}(c))}}

\\[\GAP]\>
\staterule{Plus Det}
  {\isConst{N} \qquad \densityOf{M}{E}\lam w F }
  {\densityOf{M+N}E{\lam{z}\Llet{w}{z - N\consts}F}}

\\[\GAP]\>
\staterule{Plus Rnd}
  {\isNotConst{N}\qquad\densityOf{(M,N)}{E}\lam w F}
  {\densityOf{M+N}E{\lam{z} \int\lam{w_1}\Llet{w}{(w_1,z-w_1)}F}}

\end{display}
For the \ref{Log} rule above, we require that negative arguments to $\Llog$ have zero density.
\begin{exa} \label{ex:translate}
  Letting
  $E_{7}\equiv \textsc{pdf}_{\kw{Bernoulli}(0.7)}(\kw{branch})\cdot\textsc{pdf}_{\kw{Gaussian}(0,1.0)}(\kw{temp}) \cdot\kw{isR}(\kw{branch})$,
  the sum on line 6 is evaluated using rule \ref{Plus Det} as
\[ \densityOf[\probCtx_{7}]{\kw{temp}+\kw{mB}}{E_{7}}{\lam z F_{7}} \]
where
  \[F_{7}=\Llet{\kw{temp}}{z - \kw{mB}}\int\lam{\kw{branch}} E_{7}\ \equiv
  \lambda z. 0.3\cdot\textsc{Pdf}_{\kw{Gaussian}(\kw{mB,1.0})}(z)\]
since
 $\textsc{Pdf}_{\kw{Gaussian}(0.0,1.0)}(z-r) =
 \textsc{Pdf}_{\kw{Gaussian}(r,1.0)}(z)$ for all $r$.
  Thus we obtain that the \pdf of the program fragment in Figure 1 is given by
    \[\lambda z. F_{4} + F_{7} \equiv  \lambda z.0.7\cdot\textsc{Pdf}_{\kw{Gaussian}(\kw{mA,1.0})}(z)+ 0.3\cdot\textsc{Pdf}_{\kw{Gaussian}(\kw{mB,1.0})}(z).\]
\end{exa}

Finally, as another example of compilation,
the \kw{if} statement in the program 
\begin{lstlisting}
let p = random(Beta(1.0,1.0)) in // the uniform distribution on the unit interval
let b = random(Bernoulli(p)) in
if b then p+1.0 else p
\end{lstlisting}
is handled by \ref{If Det},
yielding a density function that (modulo trivial integrals) is equivalent to 
\begin{align*}
\lambda z. \quad&\Llet{p}{z-1}\int\lambda b.[0 \le p\le 1]\cdot(\Lif{b}{p}{1-p})\cdot\kw{isL}(b) 
\\&+
\int\lambda b.[0 \le z\le 1]\cdot(\Lif{b}{z}{1-z})\cdot\kw{isR}(b)
\\
{}\equiv\lambda z. \quad&[1 \le z\le 2]\cdot(z-1) +
[0 \le z\le 1]\cdot(1-z)
\end{align*}

\subsection{Compiler Correctness}
\label{sec:compiler-correctness}

These derived judgments relate the types of 
the various terms occurring in the marg and dens judgments.
\begin{lem}[Derived Judgments]\label{lem:derivedjudgments} $ $\\
If $\wfCtx[\Gamma,\Gamma_{\probCtx}]{\probCtx}$ 
and $\dom(\Gamma_{\probCtx})=\rands\cup\dom(\consts)$
and $\Gamma,\Gamma_{\probCtx} \vdash E : \kw{real}$ 
then
\begin{enumerate}
\item
If $\marginalOf{X}{E}F$ and $X=\{x_1,\dots,x_n\}$ 
and $\Gamma_{\probCtx} \vdash (x_1,\dots,x_n) : (t_1 * \cdots * t_n)$ \\
then $\Gamma,x_1:t_1,\dots,x_n: t_n \vdash F : \kw{real}$.
\item
If $\densityOf{M}{E}{\lam z F}$ 
and $\Gamma,\Gamma_{\probCtx} \vdash M : t$ 
then $\Gamma,z:t \vdash F : \kw{real}$.
\end{enumerate}
\end{lem}
\proof$ $
  \begin{enumerate}
%
 \item By inversion of \ref{Marginal},
    $F=\int\lam{(y_1,...,y_k)} E\consts$ where   $X \subseteq \rands$ and $ (\rands \setminus X) = y_1,...,y_k$.
    By $\wfCtx[\Gamma,\Gamma_{\probCtx}]{\probCtx}$ we have that 
    $\Gamma,\Gamma_{\probCtx}\vdash\consts(x):\Gamma_{\probCtx}(x)$
    for all $x\in\dom\consts$.
    Without loss of generality, let $\Gamma_{\probCtx} \equiv \Gamma_{X},\Gamma_{Y}, \Gamma_{\consts}$ where
    $\dom{\Gamma_{X}} = X$, $\dom{\Gamma_{Y}} = \{y1,...,y_k\}$ and $\dom{\Gamma_{\consts}} = \dom\consts$.
    By repeated application of Lemma~\ref{lem:standard}.\ref{item:substitution} we obtain $\Gamma,\Gamma_{X},\Gamma_{Y} \vdash E\consts:\kw{real}$.
    By \ref{Target Int} we then derive $\Gamma,\Gamma_{X} \vdash \int \lam{(y_{1},\dots,y_{n})} E\consts:\kw{real}$.
    By repeated inversion of $\Gamma_{\probCtx} \vdash (x_1,\dots,x_n) : (t_1 * \cdots * t_n)$ we can conclude
    that ${\Gamma_{X} = x_1: t_1,...,x_n:t_n}$, which gives us the result $\Gamma,x_1:t_1,\dots,x_n: t_n \vdash F : \kw{real}$.

  \item By induction on the derivation of $\densityOf{M}{E}{\lam z F}$, using
    (1).\qed
  \end{enumerate}

The density compiler is deterministic.
\begin{lem}[Determinism]\label{lem:determinism} $ $\\
If $\wfCtx{\probCtx}$ 
and $\Gamma \vdash E : \kw{real}$ 
then
\begin{enumerate}
\item
If $\marginalOf{X}{E}{F_{1}}$
and $\marginalOf{X}{E}{F_{2}}$ then $F_{1}=F_{2}$.
\item
If $\densityOf{M}{E}{\lam z F_{1}}$ 
and $\densityOf{M}{E}{\lam z F_{2}}$ 
then $F_{1}=F_{2}$.
\end{enumerate}
\end{lem}
\proof$ $
  \begin{enumerate}
  \item The \ref{Marginal} rule is deterministic.
  \item By induction on $M$, using (1). 
    In every case, at most one compilation rule applies. \qed
  \end{enumerate}

%
We also need a technical lemma relating pure Fun expressions with their \target counterparts.
\begin{lem}\label{lem:det}
  If $M$ is pure
  and $\Gamma \vdash M : t$ and $\Gamma\vdash\sigma$ 
  then $\Pqq{M}=\Preturn \iqq M\sigma$.
\end{lem}
\proof
  By induction on $M$.
\qed

The soundness theorem asserts that, for all closed expressions $M$, the density function given by the 
density compiler indeed characterizes (via stock integration) the distribution of $M$ given by the monadic semantics.

\begin{thm}[Soundness]\label{thm:compiler-sound}
If $\densityOf[\emptyCtx]{M}1\lam z F$ 
and $\emptyCtx\vdash M:t$ 
then
\begin{align*}
(\Pqq[\;\epsilon]{M})\;A = \int_A \iqq{\lam z F} \epsilon
\end{align*}
\end{thm}
\newcommand{\LHS}{\operatorname{LHS}}
\newcommand{\RHS}{\operatorname{RHS}}
\newcommand{\Pqqsr}[1]{\Pqq[\ \rho]{{#1}\consts}}
\newcommand{\xs}{\ol{x}}
\newcommand{\ys}{\ol{y}}
\newcommand{\zs}{\ol{z}}
\newcommand{\muL}{\mu_{\mathsf{L}}}
\newcommand{\muR}{\mu_{\mathsf{R}}}
\proof
  We let $\ys:=y_1,\dots,y_n:=\rands$, and otherwise use the meta-variables from the derivation rules.

We prove the theorem by joint induction on the derivations of dens$(M)$ and $M:t$, using
the following induction hypothesis (IH):
\begin{eqnarray}
\text{ if }&&\wfCtx[\Gamma,\Gamma_{\probCtx}]{\probCtx}\text{ and } \dom(\Gamma_{\probCtx})=\rands\cup\dom(\consts)  
\\&\text{ and }&\Gamma,\Gamma_{\probCtx}\vdash M:t \text{ and } \Gamma,\Gamma_{\probCtx}\vdash E:\realT  \text{ and } \densityOf{M}E\lam z F  
\\& \text{ and }&\Gamma\vdash\rho \text{ and } \mu(B) = \int_B\iqq{\lam{\ys}E\consts}{\rho} \text{ and }\abs\mu\le1 
\\&\text{ and }&(\forall\rho' .  \Gamma_{\probCtx}\vdash\rho' \land \iqq{E}\rho\rho'\neq 0.0 \implies
    \\&&\quad ((\consts(x)=\kw{fromL}(M) \implies \exists{V}.\iqq{M\consts}\rho\rho'= \kw{inl}\;V) 
    \\&&\quad\land (\consts(x)=\kw{fromR}(M) \implies \exists{V}.\iqq{M\consts}\rho\rho'= \kw{inr}\;V))
\label{*})\\
\text{then}&&
\left(\mu\Pbind\lam{\ys}\Pqq[\ \rho]{M\consts}\right)\;A = \int_A\iqq{\lam z F}\rho.
\end{eqnarray}
\begin{FULL}

  We first note that since the density expression is only ever
  modified by multiplication with other real-valued expressions, 
  the conjunct at 3.4-3.6 of IH can only be invalidated when a deterministic variable $x=\kw{fromL}(M)$ 
  or $x=\kw{fromR}(M)$ is added to $\probCtx$,
  which only can occur in rule \ref{Match Det}.
  In the left branch of the \kw{match}, 
  $\iqq{E\cdot \kw{isL}(M\consts)}\rho\rho'\neq 0.0$ implies that $\iqq{\kw{isL}(M\consts)}\rho\rho'\neq 0.0$, so
  $\iqq{M\consts}\rho\rho'=\kw{inl}\ V$ for some $V$.
  A symmetric argument applies to the right branch of the \kw{match}.

We proceed with the induction.
  \begin{description}
  \item[\ref{Var Det}] 
    \begin{eqnarray*}
    \LHS &=& (\mu\Pbind \lambda\ys.\Preturn z\consts\rho) \;A\\
    (\text{by Lemma~\ref{lem:det}})&=& (\mu\Pbind \lambda\ys.\Pqqsr{E'}) \;A\\
    (\text{by IH})&=&\RHS
  \end{eqnarray*}
  \item[\ref{Var Rnd}] Assume that $x=y_i$, and let $\ys'=\ys\setminus x$.
\[
    \LHS = (\mu\Pbind \lambda\ys.\Preturn y_i) \;A
    =
    \int\lambda\ys.(\iqq{E\consts}\rho)\cdot[y_i\in A]
    =
    \int_A\lambda y_i.\int\iqq{\lambda\ys'.E\consts}\rho
    =\RHS
\]
  \item[\ref{Constant}] 
\[
    \LHS=(\mu\Pbind \lambda\ys.\Preturn V) \;A
    =
    \int\lambda\ys.(\iqq {E\consts}\rho)\cdot[V\in A]
    =
    (\int\iqq{\lambda\ys.E\consts}\rho)\cdot\sum_{x\in A}[x=V]
    =\RHS
\]
  \item[\ref{Fail}] $\LHS=0.0=\RHS$.
  \item[\ref{Random Const}] 
    \begin{eqnarray*}
    \LHS &=& (\mu\Pbind \lambda\ys.(\Pqqsr M\Pbind\lambda x.\mu_{\D(x)}) \;A\\
    (\text{by Lemma~\ref{lem:det}})&=& 
    (\mu\Pbind \lambda\ys.(\Preturn (\iqq{M\consts}\rho)\Pbind\lambda x.\mu_{\D(x)})) \;A\\
    (\text{by monad laws})&=& 
    (\mu\Pbind \lambda\_.\mu_{\D(\iqq{M\consts}\rho)}) \;A\\
    &=& 
    \mu_{\D(\iqq{M\consts}\rho)}(A)\cdot \int\iqq{\lambda\ys.E\consts}\rho\\
    &=&
    \int_A\iqq{\lambda z. \left(\int\lambda\ys.E\consts\right)\cdot\pdf_{\D(M\consts)}(z)}\rho=\RHS
  \end{eqnarray*}
  \item[\ref{Random Rnd}] Let $\nu\ A=\int_A\iqq{\lam w F}\rho$.
    \begin{eqnarray*}
    \LHS &=& 
    (\mu\Pbind \lambda\ys.(\Pqqsr M\Pbind\lambda z.\mu_{\D(z)})) \;A\\
    (\text{by monad laws})&=& 
    ((\mu\Pbind \lambda\ys.\Pqqsr M)\Pbind\lambda z.\mu_{\D(z)}) \;A\\
    (\text{by induction})&=& 
    (\nu\Pbind \lambda z.\mu_{\D(z)}) \;A\\
    &=& 
    \int\lambda z.\mu_{\D(z)}(A)\;d\nu\\
    &=& 
    \int\lambda z.((\iqq{\lam w F}\rho)\;z)\cdot\mu_{\D(z)}(A)\\
    &=& 
    \int\lambda w.(\iqq{F}\rho)\cdot\mu_{\D(w)}(A)\\
    &=& 
    \int\lambda w. (\iqq{F}\rho)\cdot\int_A\pdf_{\D(w)}\\
    &=& 
    \int_A\lambda z.\int\lambda w. (\iqq F\rho) \cdot\pdf_{\D(w)}(z)\\
    &=&\RHS
  \end{eqnarray*}
  \item[\ref{Tuple Var}] Let $\xs=x_{1},\dots,x_{k}$, and $\ys'=\ys\setminus\xs$. 
    Let $\nu\ A=\int_A\iqq{\lam{\xs} F}\rho$.
    \[
    \LHS = (\mu\Pbind \lambda\ys.\Preturn (\xs)) \;A
    =
    \int\lambda\ys.(\iqq{E\consts}\rho\cdot[(\xs)\in A])
    =
    \int_A\lambda(\zs).\int\iqq{\lambda\ys'.E\consts}\rho
    =\RHS
  \]
  \item[\ref{Tuple Proj L}] Let $\nu\ A=\int_A\iqq{\lam w F}\rho$.
    \begin{eqnarray*}
    \LHS &=& (\nu\Pbind \lam w \Preturn\kw{fst}\;w) \;A
    \\     &=& \int\lambda w.[\kw{fst}\ w\in A]\;d\nu
    \\     &=& \int\lambda w.[\kw{fst}\ w\in A]\cdot\iqq{F}\rho
    \\     &=& \int\lambda (z_1,z_2).\Plet{w}{(z_{1},z_{2})}{[z_{1}\in A]\cdot\iqq{F}\rho}
    \\     &=& \int_A\lambda z_{1}.\int \lambda z_{2}. \iqq{\Llet{w}{(z_{1},z_{2})}F}\rho =\RHS
    \end{eqnarray*}
  \item[\ref{Tuple Proj R}] As \ref{Tuple Proj L}.
  \item[\ref{Let Det}] 
    \begin{eqnarray*}
    \LHS &=& 
    (\mu\Pbind \lambda\ys.(\Pqqsr M\Pbind\lambda v.\Pqq[\ \rho,x\mapsto v]{N\consts})) \;A\\
    (\text{by Lemma~\ref{lem:det}})&=& 
    (\mu\Pbind \lambda\ys.(\Preturn (\iqq{M\consts}\rho)\Pbind
\lambda v.\Pqq[\ \rho,x\mapsto v]{N\consts})) \;A\\
    (\text{by monad laws})&=& 
    (\mu\Pbind \lambda\ys.\Pqq[\ \rho,x\mapsto \iqq{M\consts}\rho]{N\consts}) \;A\\ 
    &=& 
    (\mu\Pbind \lambda\ys.\Pqq[\ {\consts[\probCtx,x\mapsto M]}\rho]{N}) \;A\\ 
    &=&\RHS
  \end{eqnarray*}
  \item[\ref{Let Rnd}] Let $\nu\;B:=\int_B\iqq{\lambda\ys,x.(E\cdot F_1)\consts}\rho$.
By induction $(\Pqq[\rho'] M)\ C = \int_{C}\iqq{F_1}\rho'$ whenever $\Gamma,\Gamma_{\probCtx}\vdash\rho'$.
    \begin{eqnarray*}
    \LHS &=& 
    (\mu\Pbind \lambda\ys.(\Pqqsr M\Pbind\lambda z.\Pqq[\ \rho,x\mapsto z]{N\consts})) \;A\\
    (\text{by monad laws})&=& 
    ((\mu\Pbind \lambda\ys.(\Pqqsr M\Pbind\lambda z.\Preturn \ys,z))\Pbind\lambda \ys,z.\Pqq[\ \consts\rho,x\mapsto z]{N}) \;A
    \end{eqnarray*}
    Then
    \begin{eqnarray*}
      && (\mu\Pbind \lambda\ys.(\Pqqsr M\Pbind\lambda x.\Preturn \ys,x))\;B \\
&=& 
    \int\lambda\ys. (\Pqqsr M\Pbind\lambda x.\Preturn \ys,x)\;B\;d\mu\\
&=& 
    \int\lambda\ys.(\iqq{E\consts}\rho) \cdot(\Pqqsr M\Pbind\lambda x.\Preturn \ys,x)\;B\\
&=& 
    \int\lambda\ys.(\iqq{E\consts}\rho) \cdot\int\lambda x. [\ys,x\in B]\,d \Pqqsr M\\
    (\text{induction})&=&
    \int \lambda\ys.(\iqq{E\consts}\rho)\cdot\int\lambda x.\iqq{F_1\consts}{\rho}\cdot[\ys,x\in B]\\
    &=&
    \int_{B}\lambda\ys,x. \iqq{E\consts\cdot F_1\consts}\rho\\
    &=&
    \nu
  \end{eqnarray*}
  By induction, $(\nu\Pbind\lambda \ys,z.\Pqq[\ \rho,x\mapsto z]{N\consts}) \;A=\RHS$.
  \item[\ref{Sum Con L}] Let $\nu\ A=\int_A\iqq{\lam z F}\rho$.
\[
    \LHS = (\nu\Pbind \lam z.\Preturn\kw{inl}\;z) \;A
    = 
    \int\lambda z.[\kw{inl}\ z\in A]\;d\nu
    =
    \int_A\kw{either}\ \iqq{F}\rho\ \lambda\_.0
    =\RHS
\]
  \item[\ref{Sum Con R}] As \ref{Sum Con L}.

\item[\ref{Match Det}] Let
  \begin{align*}    
  L=\{(V_1,\dots,V_n)\mid \exists W.&
    \iqq{M\consts}\rho[y_1,\dots,y_n\mapsto V_1,\dots,V_n] = \kw{inl}\ W\\
    &\text{~and~} \emptyEnv\vdash V_i:\Gamma_{\probCtx}(y_i)\text{ for all }i\}.
  \end{align*}

  Then $\mu=\muL + \muR$ 
  where $\muL(B) = \mu(B\cap L) = \int_B\lambda(\rands).\iqq{E\consts\cdot\kw{isL}(M\consts)}\rho$
  and $\muR(B) = \mu(B\setminus L)= \int_B\lambda(\rands).\iqq{E\consts\cdot \kw{isR}(M\consts)}\rho$. 
  By additivity, we have
  \begin{eqnarray*}
    \LHS &=& (\muL + \muR\Pbind \Pqqsr{M'})\;A\\
    &=& 
    (\muL\Pbind \Pqqsr{M'})\;A + (\muR\Pbind \Pqqsr{M'})\;A
  \end{eqnarray*}
We let $E_N:=\Pmatch[z]{\iqq{M\consts}\rho}{\Pqq[{\ (\sigma,x_1\mapsto z)}]{N_1}}{\Pqq[{\ (\sigma,x_2\mapsto z)}]{N_2}}$. Then
  \begin{eqnarray*}
    (\muL\Pbind \Pqqsr{M'})\;A&=&(\muL\Pbind \lambda\ys.(\Pqqsr M\Pbind \Peither
    \\ &&\quad 
    (\lambda z. \Pqq[{\ (\sigma,x_1\mapsto z)}]{N_1})\;
    (\lambda z. \Pqq[{\ (\sigma,x_2\mapsto z)}]{N_2}))) \;A\\
    (\text{by Lemma~\ref{lem:det}})&=& 
    (\muL\Pbind \lambda\ys.(\Preturn (\iqq{M\consts}\rho)\Pbind\Peither\dots)) \;A\\
    (\text{by monad laws})&=& (\muL\Pbind \lambda\ys.E_N) \;A\\
    &=& 
    \int_L\lambda\ys.E_N(A)\;d\muL + \int_{\overline{L}}\lambda\ys.E_N(A)\;d\muL\\
    &=& 
    \int_L\lambda\ys.E_N(A)\;d\muL\\
    (\forall\ol{V}\in L\;\exists W.\; \iqq{M\consts}\rho[\ys:=\ol{V}]=\kw{inl}\ W)&=& 
    \int_L\lambda\ys. \Pqq[{\ (\rho,x_1\mapsto \kw{fromL}\ \iqq{M\consts}\rho)}]{N_1\consts} (A)\;d\muL\\
    &=& 
    (\muL\Pbind\lambda\ys. \Pqq[{\ (\rho,x_1\mapsto \kw{fromL}\ \iqq{M\consts}\rho)}]{N_1\consts})\;A\\
    (\text{by induction})&=&
    \int_A\iqq{\lam z F_1}\rho
  \end{eqnarray*}
  Symmetrically, $(\muR\Pbind \Pqqsr{M'})\;A=\int_A\iqq{\lam z F_2}\rho$, so $\LHS=\int_A\iqq{\lam z F_1}\rho + \int_A\iqq{\lam z F_2}\rho=\int_A\lam z\iqq{F_1}\rho+\iqq{F_2}\rho=\RHS$.
\item[\ref{Match Rnd}] Write $E_i := \Pqq[{\ (\rho,x_i\mapsto z)}]{N_i\consts} $. 
  Here $\Peither\;\lambda z.E_1\;\lambda z.E_2=_{\beta}$\\
  $\lambda v.\Pmatch[z]{v}{N_1}{N_2}$. 
  As in case \ref{Let Rnd} we let $\nu\;B:=
  \int_B\iqq{\lambda\ys,x_1.E\consts\cdot \Llet{w}{\kw{inl}\;x_{1}} F\consts}\rho$, and get
  \begin{eqnarray*}
    \LHS &=& (\mu\Pbind \lambda\ys.(\Pqqsr M\Pbind\lambda v.\Preturn \ys,v)
    \\ &&\quad\Pbind\lambda \ys,v.\Pmatch[z]{v}{N_1}{N_2}) \;A\\
    &=&
    \nu\Pbind\lambda \ys,v.\Pmatch[z]{v}{N_1}{N_2}) \;A \qquad (*)
  \end{eqnarray*}
  We proceed as in case \ref{Match Det} but with 
  $L:=\Set{(V_1,\dots,V_n,\kw{inl}\ W)\mid \emptyEnv\vdash V_i:\Gamma_{\probCtx}(y_i)\text{ for all }i}$, 
  yielding $(*)=\int_A\lambda z.\iqq{F_1+F_2}\rho=\RHS$.
\item[\ref{fromL}] Let $\nu\;B:=\int_B\iqq{\lam w F}\rho$.
  \begin{eqnarray*}
        \LHS &=& (\mu\Pbind \lambda\ys.\Preturn \kw{fromL}(M\consts\rho)) \;A\\
    (\text{monad law}) &=& 
    ((\mu\Pbind \lambda\ys.\Preturn M\consts\rho)\Pbind \Preturn[]\circ \kw{fromL}) \;A\\
    (\text{induction}) &=& 
    (\nu\Pbind \Preturn[]\circ \kw{fromL})\;A\\
    (\text{definition}) &=& \int_{t+u}\lam w [\kw{fromL}(w)\in A]\cdot\iqq{F}\rho
  \end{eqnarray*}
  By part 3.4-3.6 of the IH, $\iqq{F}\rho\rho'\neq0.0$ implies $\iqq{M\consts\rho\rho'}=\kw{inl}\;{V}$,
  so we have
  \[
    \int_{t+u}[\kw{fromL}(x)\in A]\cdot\iqq{F}\rho
    =
    \int_{A}\iqq{F}\rho[z:=\kw{inl}\;x] dx = \text{RHS}
  \]
\item[\ref{fromR}] As \ref{fromL}.
\item{\ref{Discrete}} Let $\nu\;B:=\int_B\iqq{\lam{\xs}F}\rho$ and 
  $\ys\setminus\xs=\zs$
  \begin{eqnarray*}
    \LHS &=& (\mu\Pbind \lambda\ys.\Preturn f(M\consts\rho)) \;A
    \\ &=&
    \int\lambda \ys.[f(\iqq{M\consts}\rho)\in A]\,d\mu\\
    &=& 
    \int\lambda \ys.[f(\iqq{M\consts}\rho)\in A]\cdot \iqq{E\consts}\rho\\
    &=& 
    \sum_{w\in A}\int\iqq{\lambda \ol{z}.[w=f(M\consts)]\cdot\int\lambda\ol{x}.E\consts}\rho
    = \RHS
  \end{eqnarray*}
\item{\ref{Plus Rnd}} Let $\nu\;B=\int_B\iqq{\lam{w}F}\rho$.
  \begin{eqnarray*}
    \LHS &=& ((\mu\Pbind \lambda\ys.\Preturn (M,N)\consts\rho)\Pbind\lam{(x_1,x_{2})}\Preturn x_{1}+x_{2}) \;A\\
    (\text{Lemma~\ref{lem:det}})&=& 
    ((\mu\Pbind \lambda\ys.\Pqqsr{(M,N)})\Pbind\lam{(x_1,x_{2})}\Preturn x_{1}+x_{2}) \;A\\
    (\text{induction})&=& 
    (\nu\Pbind\lam{(x_1,x_{2})}\Preturn x_{1}+x_{2})) \;A\\
    &=& 
    \int\lambda x,y.[x+y\in A]\,d\nu\\
    &=& 
    \int\lambda x,y.[x+y\in A]\cdot (\Plet{w}{(x,y)}{\iqq{F}\rho})\\
    (z:=x+y)&=& 
    \int_A\lambda z.\int\lambda x.\iqq{\Llet{w}{(x,z-x)}F}\rho\\
    &=& \RHS
  \end{eqnarray*}
\item{\textbf{Numeric operations on} $\realT$:} 
  Assume that $f$ is strictly monotonic 
  and $g:=f^{-1}$ has a continuous derivative 
  for $f(x)\in A$. Let $\nu\;B:=\int_B\iqq{\lam wF}\rho$.
  \begin{eqnarray*}
    \LHS &=& (\mu\Pbind \lambda\ys.\Preturn f(\iqq{M\consts}\rho)) \;A\\
    (\text{monad law}) &=& 
    ((\mu\Pbind \lambda\ys.\Preturn \iqq{M\consts}\rho)\Pbind \lam w\Preturn f(w)) \;A\\
    (\text{induction}) &=& 
    (\nu\Pbind \lam w\Preturn f(w))\;A\\
    &=& 
    \int\lam w [f(w)\in A]\cdot \iqq{F}\rho\\
    (\text{change of variables})&=& 
    \int_A\lambda z.(\iqq{\Llet{w}{g(z)}F}\rho)\cdot g'(w)=\RHS
  \end{eqnarray*}
\end{description}
  For the base case of the induction, we have that $E=1$, $\mu$ is the probability measure on the unit type, and all of $\Gamma$, $\Gamma_{\Upsilon}$, $\sigma_{\Upsilon}$ and $\rho$ are empty. Clearly, (IH) holds for the base case.
\qed
\end{FULL}

Part 3.4-3.6 of the induction hypothesis above is used when attempting
to evaluate match-bound variables (e.g., $x=\kw{fromL}(M)$) for valuations that give the other
branch (e.g., $\iqq{M}\sigma = \kw{inr}\;V$).  
For such valuations the density is always zero (since, e.g., $\kw{isL}(\kw{inr}\;V) = 0.0$).

\section{Evaluation}
\label{sec:evaluation}
\newcommand{\orig}{orig}
\newcommand{\loc}{\textsc{loc}\xspace}

We evaluate the compiler on
several synthetic textbook examples 
and several real examples from scientific applications.
We wish to validate that the density compiler handles these examples, 
and understand how much the compiler reduces the developer burden, 
and its performance impact.  

\subsection{Implementation}
Since \fun is a sublanguage of \fsharp, we implement our models as \fsharp programs,
and use the quotation mechanism of \fsharp to capture their syntax trees.
Running the \fsharp program corresponds to sampling data from the model.
To compute the \pdf, the compiler takes the syntax tree (of \fsharp type \kw{Expr}) of the model
and produces another \kw{Expr} corresponding to a deterministic \fsharp program as output.  
We then use run-time code generation to compile the generated \kw{Expr} to MSIL bytecode, 
which is just-in-time compiled to executable machine code when called, 
just as for statically compiled \fsharp code.  
%
Our implementation supports immutable arrays and records, 
which are both translated using adaptations of the corresponding rules for tuples.
For efficiency, the implementation must avoid introducing redundant computations,
translating the use of substitution in the formal rules to more efficient \kw{let}-bindings that
share the values of expressions that would otherwise be
re-computed.
As is common practice, our implementation and Filzbach \citep{Filzbach} both work with
the \emph{logarithm} of the density, which avoids products of densities
in favor of sums of log-densities where possible, to avoid numerical underflow.
It also performs some simple but effective peephole optimization to elide
canceling applications of \lstinline{Log} and \lstinline{Exp} and additions of~$0$. 

\subsection*{Code Examples}
\begin{exa}[Mixture Of Gaussians]\label{exa:mog}

To illustrate the implementation, here is the actual \fsharp code expressing a mixture of Gaussians (a variant of our introductory example):
\begin{lstlisting}
  type W = {bias: double; mean: double[]; sd: double[]}

  [<Fun>]
  let prior () = 
   { bias = random(Uniform(0.0, 1.0))
     mean = [| for i in 0..1 -> random(Uniform(-1000.0, 1000.0)) |]
     sd   = [| for i in 0..1 -> random(Uniform(100.0, 500.0)) |] }
   
  let xs = [| for i in 1..100 -> () |]

  [<Fun>]
  let model w = [| for x in xs -> 
                   if random(Bernoulli(w.bias))
                   then random(Gaussian(w.mean.[0],w.sd.[0])
                   else random(Gaussian(w.mean.[1],w.sd.[1]) |]
\end{lstlisting}
The code uses both records (to structure the prior \lstinline{w} of record type \lstinline{W}) and arrays;
both are encoded as tuples in the core language.
The model function receives multiple inputs in an array \lstinline{xs} and return an array multiple outputs from the model. The outputs of \lstinline{model} are constructed using an \emph{array comprehension}. The \lstinline{[<Fun>]} attributes declare that definitions
\lstinline{prior} and \lstinline{model} should be made available as quoted expression trees (as well as executable functions) so their code can be inspected by
the density compiler.  

\clearpage
The probability density function compiled for function \lstinline{model} is (after manual reformatting to match the notation in this article): 

\begin{lstlisting}
let logPdf = 
  fun w (ys:real[]) ->
    logprodBy((fun i ->
        let x = xs.[i]
        Log(Exp(Log(pdf_Gaussian(w.mean.[0], w.sd.[0], ys.[i])) +
                let b=true in Log(pdf_Bernoulli(w.bias,b)))
            +
            Exp(Log(pdf_Gaussian(w.mean.[1], w.sd.[1], ys.[i])) +
                let b=false in Log(pdf_Gaussian(w.bias, b))))),
        xs.Length)
\end{lstlisting}
The helper function \lstinline{logprodBy($f,n$)} computes the sum of the log densities $f(0) + \cdots + f(n-1)$.
Notice the insertion of logarithms to avoid underflow.

The effect of disabling our simple peephole optimizer is to produce both less readable and less efficient code:
\begin{lstlisting}
let logPdf = 
  fun w (ys:real[]) ->
    logprodBy(
       (fun i ->
        let x = xs.[i]
        Log(Exp(Log(pdf_Gaussian(w.mean.[0],w.sd.[0], ys.[i]))+
                Log(Exp(Log(1.0)+
                        let b2=true
                        Log(pdf_Bernoulli(w.bias, b2))+
                        Log(Exp(Log(1.0))))))+
            Exp(Log(pdf_Gaussian(w.mean.[1],w.sd.[1], ys.[i1]))+
                Log(Exp(Log(1.0)+
                        let b2=false
                        Log(pdf_Bernoulli(w.bias, b2))+
                        Log(Exp(Log(1.0)))))))),
       xs.Length)
\end{lstlisting}
\end{exa}

\begin{exa}[Linear Regression]\label{exa:lr}

For an example involving arithmetic we take {\it linear regression}.
Given some noisy sample of points, the task is to estimate the parameters of a line fitting the points, yielding the line's slope \lstinline{a},  intercept \lstinline{b} and an estimate of the \lstinline{noise}.

The generative model is expressed as the following \fsharp code:

\begin{lstlisting}
  type W = {a: double; b: double; noise: double}

  [<Fun>]
  let prior () = 
    { a = random(Uniform(-1000.0, 1000.0)) 
      b = random(Uniform(-1000.0, 1000.0))
      noise = random(Uniform(0.001, 100.0))}

  let xs = [| -100.0 .. 100.0 |]

  [<Fun>]
  let model w = [| for x in xs -> 
                      let m =  w.a * x + w.b 
                      let d = w.noise
                      random(Gaussian(m, d)) |]  
\end{lstlisting}

The (log) probability density function compiled for function \lstinline{model} is: 
\begin{lstlisting}
  let logPdf =
    fun w ys ->
      logprodBy((fun i ->
                    let x = xs.[i]
                    Log(pdf_Gaussian(w.a*x+w.b, w.noise, ys.[i]))),
                xs.Length) 
\end{lstlisting}
\end{exa}

\begin{exa}[Mixture Of Regressions]\label{exa:mlr}

Combining aspects of the previous two examples we construct a mixture of two linear regressions, in which the slope and intercept of the line
is selected by a latent boolean indicator variable.

\begin{lstlisting}
  type W = {bias:double; a: double[]; b: double[]; noise: double}

  [<Fun>]
  let prior () = 
    { bias = rand.Uniform(0.0, 1.0)
      a = [| for i in 0..1 -> rand.Uniform(-1000.0, 1000.0)|]  
      b = [| for i in 0..1 -> rand.Uniform(-1000.0, 1000.0)|] 
      noise = rand.Uniform(0.001, 100.0) }

  let xs = [| -100.0 .. 100.0 |]

  [<Fun>]
  let model w =
    [| for x in xs -> 
           if rand.Bernoulli(w.Bias)
           then let m = w.a.[0] * x + w.b.[0]
                rand.Gaussian(m, w.noise)
           else let m = w.a.[1] * x + w.b.[1]
                rand.Gaussian(m,  w.noise)|]              
\end{lstlisting}

\clearpage
The (log) probability density function compiled for function \lstinline{model} is: 
\begin{lstlisting}
  let logPdf =
    fun w ys ->
    logprodBy(
          (fun i ->
            let x = xs.[i]
            Log(
             Exp(Log(pdf_Gaussian(w.a.[0]*x+w.b.[0], w.noise, ys.[i])) +
                 let b=true in Log(pdf_Bernoulli(w.bias, b)))
             +
             Exp(Log(pdf_Gaussian(w.a.[1]*x+w.b.[1], w.noise, ys.[i])) +
                 let b=false in Log(pdf_Bernoulli(w.bias, b))))),
         xs.Length);
\end{lstlisting}
\end{exa}

\subsection{Metrics}

We consider scientific models 
with existing implementations for \MCMC-based inference, written by domain experts. 
We are interested in how the modelling and inference experience would change,
in terms of developer effort and performance impact,
when adopting the \fun-based solution.

We assess the reduction in developer burden
by measuring the code sizes (in lines-of-code (\loc)) 
of the original implementations of model and density code, 
and of the corresponding \fun model.  
For the synthetic examples, we have written both the model and the density code.
The original implementations of the scientific models contain 
helper code such as I/O code for reading and writing data files in an application-specific format.
Our \loc counts do not consider such helper code, but only count 
the code for generating synthetic data from the model, 
code for computing the logarithm of the posterior density of the model, 
and model-related code for setting up and interacting with Filzbach itself.  
We also compare the running times 
of the original implementations versus the \fun versions
for \MCMC-based inference using Filzbach, 
not including data manipulation before and after running inference.



\begin{table}[t]
\small
\begin{center}
\begin{tabular}{|l|c|c|c|c|c|c|c|}
\hline
\textbf{Example} 
& \textbf{\orig }
& \textbf{\loc, \orig }
& \textbf{\loc, \fun }
& 
& \textbf{time (s), \orig }
& \textbf{time (s), \fun }
& \\
\hline
mixture of Gaussians & \fsharp & 32 & 20 & 0.63x & 0.74 & 1.28 & 1.7x \\
\hline
linear regression & \fsharp & 27 & 18 & 0.67x & 0.21 & 0.55 & 2.6x \\
\hline
mixture of regressions & \fsharp & 43 & 28 & 0.65x & 1.02 & 2.09 & 2.0x \\
\hline
species distribution & C\# & 173 & 37 & 0.21x & 36 & 67 & 1.9x \\
\hline
net primary productivity & C\# & 82 & 39 & 0.48x & 3.7&  6.1 & 1.6x \\
\hline
global carbon cycle & C\# & 1532 & 402 & 0.26x & n/a & 301 & n/a \\
\hline
\end{tabular}
\end{center}
\caption{Lines-of-code and running time comparisons of synthetic and scientific models.}
\label{tab:results}
\end{table}

\paragraph{\textbf{Synthetic examples}}
Our synthetic examples are 
three classic problems:
the unsupervised learning task \emph{mixture of Gaussians} (Example \ref{exa:mog})
the supervised learning task \emph{linear regression} (Example \ref{exa:lr}), 
and a \emph{mixture of regressions} (Example \ref{exa:mlr}).
Example \ref{exa:mog} can be thought of as a probabilistic version of \emph{$k$-means clustering}: inference is trying to determine 
the unknown mixing bias and the means and variances of the Gaussian components.  
In Example \ref{exa:lr} 
inference is trying to determine the coefficients of the line.
In Example \ref{exa:mlr} 
inference is trying to determine the coefficients and mixing bias of two lines.

\paragraph{\textbf{Species distribution}~\citep{mcinerny2011fine}}  
The species distribution problem is to give the probability that 
certain species will be present at a given site, based on climate factors.  
It is a problem of long-standing interest in ecology and has
taken on new relevance in light of the issue of climate change.  
The particular model that we consider  
is designed to mitigate \emph{regression dilution} 
arising from uncertainty in the predictor variables, 
for example, measurement error in temperature data.  
Inference tries to determine various features of the species and the environment,
such as the optimal temperature preferred by a species, 
or the true temperature at a site (see code and density function in Appendix~\ref{app:sd}).


\paragraph{\textbf{Global carbon cycle.}~\citep{smith2012carbon}}
The dynamics of the Earth's climate are intertwined with the
terrestrial carbon cycle, and better carbon models (modelling how
carbon in the air gets converted to biomass) enable better constrained
projections about these systems.  We consider a fully
data-constrained terrestrial carbon model,  
which is composed of various submodels for smaller processes such as 
\emph{net primary productivity}, the fine root mortality rate, 
or the fraction of trees that are evergreen versus deciduous.  
Inference tries to determine the different parameters of these submodels.

\paragraph{\textbf{Discussion.}}
Table~\ref{tab:results} reports the metrics for each example.
The \loc numbers show significant reduction in code size, with more
significant savings as the size of the model grows.  The larger models
(where the \fun versions are $\approx 25\%$ of the size of the original) 
are more indicative of the savings in developer and maintenance effort
because Filzbach interaction code, which is roughly the same in all models, takes up a larger fraction of the smaller models.
We find the running times encouraging: we have made little attempt to
optimize the generated code, and preliminary testing indicates that
much of the performance slow-down is due to constant factors.

The global carbon cycle model is composed of submodels, each with their own
dataset.  Unfortunately, it is unclear from the original source code
how this composition translates to a run of inference, making it
difficult to know what constitutes a fair comparison.  
Thus, we do not report a running time for the full model.  
However, we can measure the running time of individual submodels, such
as net primary productivity, where the data and control flow are simpler.

\section{Related Work}
\label{sec:related}

We have presented the first algorithm for deriving density functions from generative processes,
that is both proved correct and implemented.
An abridged version of this paper appears as~\cite{tacas13:densities}.
The correctness proof 
(for a version of the source language without pure \kw{let} and general \kw{match}) 
was recently mechanized in Isabelle by~\cite{DBLP:conf/esop/EberlHN15}.

This paper builds on work by 
\cite{DBLP:conf/popl/BhatAVG12}
who develop a theoretical framework for computing \pdf{s}, 
but describe no implementation nor correctness proof.
The density compiler of Section~\ref{sec:density-compiler}
has a simpler presentation, with two judgments compared to five,
and has rules for pure \kw{let}s and operations on integers.
Our paper also uses a richer language (\fun), 
which adds \kw{fail}, \kw{match} and general \kw{if} 
(and for performance reasons, pure \kw{let}).

\cite{modelLearner} describe a naive density calculation routine 
for \fun without random \kw{let}s; 
this sublanguage does not cover many useful classes of models 
such as hierarchical and mixture models.

The BUGS system computes densities from declaratively specified models
to perform Gibbs sampling \citep{GTS94:Bugs}.  However, the models are
not compositional as in this work, and only the joint density over all
variables is possible.  
%
%
The AutoBayes system also computes densities for deriving maximum
likelihood and Bayesian estimators for a significant class of
statistical models~\citep{Schumann08:AutoBayes}.  It is not formally
specified and does not appear to be compositional.  Neither system
addresses the non-existence of \pdf{}s, presumably restricting
expressivity in order to avoid the issue.

Stan~\citep{stan-software:2014} is a probabilistic programming language that supports various forms of \MCMC.
The Stan language is a derivative of BUGS, and is compiled to efficient C++ sampling code, in part based on automatically derived density functions.
For models with latent variables, 
including mixture models such as mixture of Gaussians,
and for models that perform non-linear computations on random values,
Stan's users are required to manually manipulate the log probability density
function, using a primitive operation \texttt{increment\_log\_prob}.
Stan employs \emph{automatic differentiation}~\citep{griewank08:AutomaticDifferentiation} 
of log posteriors in order to apply gradient-based Hamiltonian \MCMC algorithms.  
This relieves the user from coding error-prone derivatives.


Inference for the Church language also uses \MCMC, 
but works with distributions over runs of a program 
instead of over its return value~\citep{wingate2011lightweight}, 
circumventing the need for a \pdf.

There are many other systems for probabilistic programming, 
some of which provide a way to compute density functions or an analogous object; 
however, they sacrifice some other feature to do so.  
Several languages only provide support for finite, discrete distributions, 
but provide access to the probability mass function~\citep{DBLP:conf/popl/RamseyP02,KS09:EmbeddedProbabilisticProgramming}.
Like BUGS and Stan, systems like the Hierarchical Bayes Compiler~\citep{hbc} 
are not formally defined and require models to be specified in a monolithic way, 
whereas large models in Fun can be composed of smaller models.  
Probabilistic logic languages like Markov Logic~\citep{Domingos:2008:ML:1793956.1793962} 
do not have generative semantics 
but instead have semantics in \emph{undirected graphical models} 
which are equipped with \emph{potential functions}
that are analogous to density functions.  
The language is constructed in a way that guarantees the existence of a potential function, 
but which eliminates the possibility to express models that require pure \kw{let},
such as the global carbon cycle model.

\section{Conclusions and Future Work}
\label{sec:conclusions}
We have described a compiler for 
automatically computing probability density functions for programs from 
a rich Bayesian probabilistic programming language, 
proven the algorithm correct, 
and shown its applicability to real-world scientific models.

The inclusion of \kw{fail} in the language 
appears useful for scientific models, 
giving a simple facility to exclude branches 
that are scientifically impossible from consideration.
However, more investigation is needed to settle this claim.

A drawback of the compiler is that terms of composite type 
are required either to have a \pdf
or to be pure, ruling out terms such as 
\kw{(0.0, random(Uniform))}.
One possibility for future work would be to refine the types of expressions
with determinacy information, 
and make use of this additional information to admit more joint distributions (cf.~\ref{Tuple Var}).

\section*{Acknowledgments}
We thank Manuel Eberl and Tobias Nipkow for helpful comments,
in particular on the semantics of the integration operator.

\appendix

\section{Species Distribution~\citep{mcinerny2011fine}}\label{app:sd}

\begin{lstlisting}
let Nspecies = 20
let Nsamples = 2000

type W = 
 { Topt:     double[] 
   Tbreadth: double[] 
   MaxProb:  double[] 
   Terr:     double
   Yerr:     double
   Ttrue:    double[]}

type Y = {Tobs: double[]; Y: double[][]} 

let CalcSpProb w t sp =
  let z = (t - w.Topt.[sp]) / w.Tbreadth.[sp]
  w.MaxProb.[sp] * exp (- z*z)

[<Fun>]
let prior () = 
  { Topt     = [| for j in 0..Nspecies-1 -> random(Uniform(0.1, 50.0)) |]
    Tbreadth = [| for j in 0..Nspecies-1 -> random(Uniform(0.1, 50.0)) |]
    MaxProb  = [| for j in 0..Nspecies-1 -> random(Uniform(0.1, 1.0)) |]
    Terr     = random(Uniform(0.1, 10.0))
    Yerr     = random(Uniform(0.01, 0.5))
    Ttrue    = [| for i in 0..Nsamples-1 -> random(Uniform(5.0, 30.0))|] }

let samplesR = [|0..Nsamples-1|]
let speciesR = [|0..Nspecies-1|]
 
[<Fun>]
let model w =
  let tobs = [| for i in samplesR -> random(Gaussian(w.Ttrue.[i], w.Terr)) |]
  let y =  [| for i in samplesR -> 
              [| for j in speciesR -> 
                 let p = CalcSpProb w w.Ttrue.[i] j
                 random(Gaussian(p, w.Yerr)) |] |] 
  { Tobs = tobs
    Y = y }
\end{lstlisting}

\begin{lstlisting}
// the generated log probability density function for model
let logPdf =
  fun w ys ->
    let y = ys.Y 
    let tobs = ys.Tobs
    let i0 = samplesR
    (logprodBy((fun i1->
                let i=samplesR.[i1]
                Log(pdf_Gaussian(w.Ttrue.[i],w.Terr,tobs.[i1]))),
               samplesR.Length)) +
    (let i2 = samplesR
     logprodBy((fun i3->
                let i=samplesR.[i3]
                let i4=speciesR
                logprodBy((fun i5->
                           let j=speciesR.[i5]
                           Log(pdf_Gaussian(CalcSpProb w (w.Ttrue.[i]) j,w.Yerr,y.[i3].[i5]))),
                          speciesR.Length)),
               samplesR.Length))
\end{lstlisting}


\end{document}